\documentclass[12pt,twoside]{iopart}

\usepackage[bottom]{footmisc} 
\DefineFNsymbols{rjnumber}{{*}{^1}{^2}{^3}{^4}{^5}{^6}{^7}{^8}{^9}
{^{10}}{^{11}}{^{12}}{^{13}}{^{14}}{^{15}}{^{16}}{^{17}}{^{18}}{^{19}}
{^{20}}{^{21}}{^{22}}{^{23}}{^{24}}{^{25}}{^{26}}{^{27}}{^{28}}{^{29}}
{^{30}}{^{31}}{^{32}}{^{33}}{^{34}}{^{35}}{^{36}}{^{37}}{^{38}}{^{39}}{^{40}}} 
\setfnsymbol{rjnumber} 
\usepackage{psfrag}
\usepackage{epsfig}
\newcommand{\eq}[1]{(\ref{#1})}
\newcommand{\fat}[1]{\mbox{\boldmath$#1$}}
\newcommand{\fig}[1]{figure \ref{#1}}

\newcommand{\script}[1]{{\textrm{\scriptsize #1}}}
\newcommand{\su}[1]{_\script{#1}}
\newcommand{\suj}[1]{_{\textrm{\scriptsize #1}}}	
\newcommand{\sut}[1]{_{\textrm{\tiny #1}}}
         
\begin{document}
\title{Inertial forces and the foundations of optical geometry}%
\author{Rickard Jonsson\\[2mm]%
{\small \it Department of Theoretical Physics,}\\
{\small \it Chalmers University of Technology, 41296 G\"oteborg,
Sweden}
\\[2mm]
\small{\rm E-mail: rico@fy.chalmers.se}
\\[2mm]
\small{\rm
Submitted 2004-12-10, Published 2005-12-08\\
Journal Reference: Class. Quantum Grav. {\bf 23} 1
}
}
\begin{abstract}
Assuming a general timelike congruence of worldlines as a reference frame, 
we derive a covariant general formalism of inertial forces in General
Relativity.  Inspired by the works of Abramowicz et. al. (see e.g.
Abramowicz and Lasota 1997 {\it Class. Quantum Grav.} {\bf 14} A23-30), we also study conformal rescalings of spacetime and
investigate how these affect the inertial force formalism. 
While many ways of describing spatial curvature of a trajectory has
been discussed in papers prior to this, one particular prescription 
(which differs from the standard projected curvature 
when the reference is shearing) appears novel.   
For the particular case of a hypersurface-forming congruence, using a
suitable rescaling of spacetime, we show that a geodesic photon is always
following a line that is spatially straight with respect to the new
curvature measure. This fact is intimately
connected to Fermat's principle, and allows for a
certain generalization of the optical geometry as will be further
pursued in a companion paper (Jonsson and Westman 2006 {\it Class. Quantum Grav.} {\bf 23} 61). 
For the particular case when the shear-tensor vanishes, we present
the inertial force equation in three-dimensional form
(using the bold face vector notation), and note how similar it is to
its Newtonian counterpart. 
From the spatial curvature measures that we introduce, we derive
corresponding covariant differentiations of a vector defined along a
spacetime trajectory. This allows us to connect the formalism of this
paper to that of Jantzen et. al. (see e.g. Bini et. al. 1997 {\it Int. J. Mod. Phys. D} {\bf 6} 143-98).
\end{abstract}

\section{Introduction}
Inertial forces, such as centrifugal and
Coriolis forces, have proven to be helpful in Newtonian mechanics.
Quite a lot of attention has been given to generalizing the concept to
General Relativity. In fact the last fifteen years there has been
a hundred or so papers related to inertial forces in General
Relativity. For an overview see \cite{jantzen1}.

Many of these articles are related to particular types of spacetimes, and
special types of motion. There are also a few that are completely
general. This article is of the latter kind. 
The scope is to develop a covariant formalism,  applicable to any
spacetime, and any motion of a test particle, using an arbitrary reference congruence of timelike worldlines. 
In view of the already existing bulk of papers we will keep the
introductory remarks to a minimum here and just outline the contents
of the article.

In section 2 we introduce the basic notation of the article. 

In section 3 we 
derive a spatial curvature measure for a spacetime
trajectory. We do this by projecting the trajectory down along the reference
congruence onto the local time slice. We also derive how the time
derivative of the speed relative to the congruence is related to the
four-acceleration of the test particle. The resulting equations we
put together to form a single equation that relates the test particle
four-acceleration (and four-velocity) to the spatial curvature, the
time derivative of the speed and the local derivatives of the 
congruence four-velocity. The terms connected to the congruence
derivatives can be regarded as inertial forces.  We also express the four-acceleration of the
particle in terms of the experiences comoving forces, as well as in
terms of the forces as {\it given} by the congruence observers.

In section 4 we introduce a different kind of spatial curvature
measure. The new curvature measure is such that when we are following
a straight line with respect to this measure, the spatial distance
traveled (as defined by the congruence) is minimized (with respect to
variations in the spatial curvature).  This is in fact not the case
for the standard projected curvature when the congruence is
shearing. Using the new curvature measure we create a slightly different inertial
force formalism. 

In section 5 we consider general conformal rescalings of spacetime,
and how these affect the inertial force formalism.

In section 6 we consider a foliation of spacetime into spacelike
time slices and a corresponding orthogonal congruence. Given a labeling
$t$ of the time slices we rescale away time dilation with respect to
$t$. Relating spatial curvature etc to the rescaled spacetime, but
considering the real (non-rescaled) forces, we find an inertial force
formalism that is very similar to the already derived formalisms of
this paper.
We show that a geodesic photon always follows a straight line in the sense of
section \ref{news}. We also show that it follows a straight line in the
projected sense if the congruence is shearfree.
These results allows certain generalizations of the
optical geometry (for an introduction to optical geometry see e.g. \cite{optiskintro}) as will be pursued in a 
companion paper \cite{genopt}.

In section 7 we show that the
fact that a geodesic photon follows a straight line
in the new sense relative to the rescaled spacetime follows
from Fermat's principle. 

In section 8 we introduce two new curvature measures related to geodesic photons,
and what we {\it see} as straight, and use these in the inertial force
formalism.

In section 9 we summarize the inertial force formalisms (excepting
those related to rescalings) connected to
the various introduced curvature measures.

In section 10 we rewrite the four-covariant formalism as a
three-dimensional formalism, for the particular case of vanishing
shear (assuming only isotropic expansion). While fully
relativistically correct, in this form the inertial force formalism is
very similar to its Newtonian counterpart. 

In section 11 we derive a spacetime transport law of a vector,
corresponding to {\it spatial} parallel transport with respect to the
spatial geometry defined by the reference congruence. 

In section 12 we consider an alternative approach to inertial forces
resting on the transport equation of section 11.

In section 13 we connect to the approach of Jantzen et. al. 

In section 14 we conclude the article. Then follows the appendixes.

\section{The basic notation}\label{basic}
In a general spacetime, we consider an arbitrary reference congruence
of timelike worldlines of four-velocity $\eta^\mu$. Each such worldline
corresponds to events at a single spatial point in our frame of
reference. 
We can split the four velocity $v^\mu$ of a test particle into a part
parallel to $\eta^\mu$ and a part orthogonal to $\eta^\mu$:
\begin{eqnarray}\label{hit}
v^\mu=\gamma(\eta^\mu + v t^\mu).
\end{eqnarray}
Here $v$ is the speed of the test particle relative to the congruence
and $\gamma$ is the corresponding $\gamma$-factor. The vector $t^\mu$ is a
normalized spatial vector 
(henceforth vectors that are orthogonal to $\eta^\mu$ will be referred to as {\it spatial} vectors), 
pointing in the (spatial) direction of motion.

Projected spatial curvature and direction of curvature we will denote
by $R$ and $n^\mu$, the latter being a normalized spatial vector. By projected
curvature we
mean that we project the spacetime trajectory in question down along the
congruence onto the local slice%
\footnote{If the congruence has no rotation there exists a finite
sized slicing orthogonal to the congruence. If the congruence is
rotating we can still introduce a slicing that is orthogonal at the
point in question. It is easy to realize that whatever such locally
orthogonal slicing we
choose, the projected curvature and curvature directions will be the same, and are thus well defined.}
and evaluate the spatial curvature there.
There are also several alternative definitions of curvature and
curvature direction. In particular we will us $\bar{R}$ and
$\bar{n}^\mu$ to denote what we will call the 'new-straight' curvature
and curvature direction, to be introduced in section \ref{news}. 

Throughout the article we will use $c=1$ and adopt the spatial sign convention
$(-,+,+,+)$. The projection operator\footnote{Applying this tensor to a vector extracts the spatial (i.e. orthogonal to $\eta^\mu$) part of the vector.} 
along the congruence  then takes the form ${P^\alpha}_\beta\equiv
{g^\alpha}_\beta+\eta^\alpha \eta_\beta$. We also find it
convenient to introduce the suffix $\perp$. When applied to a
four-vector, as in $[K^\mu]_\perp$, it selects
the part within the brackets that is perpendicular to both $\eta^\mu$
and $t^\mu$. 

\section{Inertial forces using the projected curvature}\label{inett}
The objective with this section is to go from the spacetime equations of
motion for a test particle and derive an expression for $R$, $n^\mu$ and the time
derivative of $v$, in terms of $v$ and $t^\mu$ for given forces and congruence behavior. 

\subsection{The projected curvature and curvature direction}
\begin{figure}[ht]
  \begin{center}
    \psfrag{t}{$t$}
      	\epsfig{figure=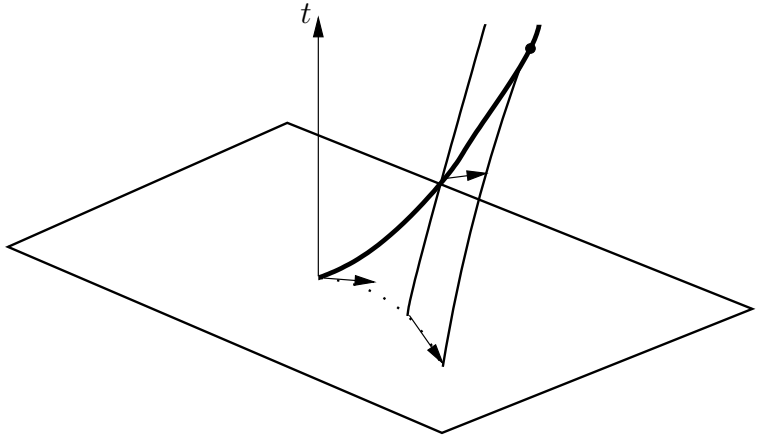,width=10cm,angle=0}
	\caption{A 2+1 illustration of a projection of a spacetime
      	trajectory onto a time slice, seen from freely falling
      	coordinates, locally comoving with the reference congruence. 
	}  
     	\label{twister}
  \end{center} 
\end{figure}

The idea behind the projected curvature with respect to the congruence is
illustrated by \fig{twister}. Notice that the time-slice we are depicting is
only assumed to be orthogonal to the congruence at the point where the
test particle worldline intersects the slice.

Taking the covariant derivative
$\frac{D}{D\tau}$ along the test particle worldline, of \eq{hit} we
readily find
\begin{eqnarray}\label{hrrr}
\left[ \frac{D v^\mu}{D\tau}  \right]_\perp = 
\gamma^2 [a^\mu]_\perp+\gamma^2 v [t^\alpha \nabla_\alpha
  \eta^\mu]_\perp + \gamma v \left[ \frac{D t^\mu}{D\tau}
  \right]_\perp.
\end{eqnarray}
Here $a^\mu$ is the four-acceleration of the congruence.
Now we want to relate the covariant derivative of $t^\mu$ in \eq{hrrr} to the
projected curvature. As concerns the $\perp$-part of this we can
consider the covariant derivative to stem from a two-step process. First
we transport it along the curved projected trajectory, then we
Lie-transport it up along the congruence as depicted in \fig{twister}%
\footnote{Letting $ds=v d\tau_0=v\gamma d\tau$, we have in freely
  falling coordinates $[dt^\mu]_\perp=\frac{n^\mu}{R}ds + 
[t^\alpha \nabla_\alpha \eta^\alpha]_\perp d\tau_0$ from which \eq{ff} follows immediately.}. 

Alternatively, we may in the style of \cite{foertsch}, consider the worldsheet spanned by
the congruence lines that are crossed by the test particle
worldline. On this sheet we can uniquely extend the forward vector
$t^\mu$, defined along the test particle worldline, into a vector
field that is tangent to the sheet, normalized and orthogonal to $\eta^\mu$.
Considering an arbitrary smooth extension of this field $t^\mu$ around
the sheet, the projected curvature can be written as
$\frac{n^\mu}{R} =[t^\alpha \nabla_\alpha t^\mu]_\perp$. We also realize that,
as concerns the $\perp$-part, this field will be Lie-transported into
itself (in the $\eta^\mu$ direction). Thus we have 
$[\eta^\alpha \nabla_\alpha t^\mu]_\perp= [t^\alpha \nabla_\alpha
  \eta^\mu]_\perp$. Then we we can write
\begin{eqnarray}
\left[ \frac{D t^\mu}{D\tau}  \right]_\perp &=& 
[\gamma (\eta^\alpha + vt^\alpha) \nabla_\alpha t^\mu]_\perp \\
&=&\gamma [t^\alpha \nabla_\alpha \eta^\mu]_\perp + \gamma v
\frac{n^\mu}{R} \label{ff}
\end{eqnarray}
Using this together with \eq{hrrr} we get
\begin{eqnarray}\label{perpdone}
\frac{1}{\gamma^2}\left[\frac{D v^\mu}{D\tau}\right]_\perp=
\left[a^\mu \right]_\perp + 2v \left[t^\alpha \nabla_\alpha \eta^\mu
\right]_\perp + v^2 \frac{n^\mu}{R}.
\end{eqnarray}
So here is a general contravariant expression for the local projected
curvature of a spacetime trajectory.

\subsection{The speed change per unit time}
Now we would like a corresponding expression for the speed
change per unit time.  We have
$\gamma=-v^\alpha \eta_\alpha$. Differentiating both sides of this
expression with respect to the proper time $\tau$ along the trajectory readily yields
\begin{eqnarray}
\gamma^3 v \frac{dv}{d\tau}&=&-\frac{Dv^\alpha}{D\tau} \eta_\alpha -
\frac{D\eta^\alpha}{D\tau} v_\alpha  \\
&=&-\frac{Dv^\alpha}{D\tau} \left(\frac{v_\alpha}{\gamma} - v
t_\alpha\right)- \frac{D\eta^\alpha}{D\tau} \gamma(\eta_\alpha+v
t_\alpha)\\
&=&v t_\alpha\frac{Dv^\alpha}{D\tau}- \frac{D\eta^\alpha}{D\tau}
\gamma v t_\alpha \label{fo}.
\end{eqnarray}
In the last equality we used the normalization of $\eta^\mu$ and
$v^\mu$. Notice that the differentiation is along the trajectory in
question, so we have
\begin{eqnarray}\label{kattaett}
\frac{D\eta^\alpha}{D\tau}	  
&=&\gamma (\eta^\rho +  v t^\rho)  \nabla_\rho \eta^\alpha.
\end{eqnarray}
Using this in \eq{fo} we readily find
\begin{eqnarray}\label{finaldv}
\frac{1}{\gamma^2}\frac{Dv^\alpha}{D\tau} t_\alpha&=& 
t_\alpha (\eta^\rho + v t^\rho) \nabla_\rho \eta^\alpha + \gamma \frac{dv}{d\tau}.
\end{eqnarray} 
So here we have a covariant equation for the speed change as well.

\subsection{Putting it together}
Multiplying \eq{finaldv} by $t^\mu$ and adding it to
\eq{perpdone}, we get a single vector equation that relates the
four-acceleration to both the speed change and the projected spatial
curvature
\begin{eqnarray}
\frac{1}{\gamma^2}\left( \left[\frac{Dv^\mu}{D\tau}\right]_\perp +
 \frac{Dv^\alpha}{D\tau} t_\alpha t^\mu \right)=&&{\hspace{-0mm}} \left[a^\mu \right]_\perp
+ 2v \left[t^\alpha \nabla_\alpha \eta^\mu \right]_\perp + v^2
 \frac{n^\mu}{R} + \nonumber\\&&{\hspace{-0mm}}
 t^\mu t_\alpha(\eta^\rho + v t^\rho)\nabla_\rho \eta^\alpha  + \gamma
\frac{dv}{d\tau} t^\mu .
\end{eqnarray}
This can be simplified to
\begin{eqnarray}\label{rattok}
\frac{1}{\gamma^2}{P^\mu}_\alpha \frac{Dv^\alpha}{D\tau}
= a^\mu 
+ 2v \left[t^\alpha \nabla_\alpha \eta^\mu \right]_\perp +  v t^\mu t^\alpha
  t^\rho\nabla_\rho \eta_\alpha  + \gamma
\frac{dv}{d\tau} t^\mu + v^2 \frac{n^\mu}{R}.
\end{eqnarray}
So here we have a generally covariant relation between the
four-acceleration, the projected curvature and the speed change.

\subsection{Experienced  forces and the kinematical invariants}
To make it more clear what an observer performing the specified motion 
actually experiences, we can rewrite the left hand side of \eq{rattok}
in terms of the experienced forward thrust%
\footnote{By definition the observers forward direction is the
direction from which he sees the congruence points coming (assuming he
has some way of seeing them).}
$F_\parallel$ and the
experienced
 sideways thrust $F_\perp$. This is a simple
exercise of special relativity performed in 
\ref{app_specialforce}. We may then write
\begin{eqnarray}\label{kin2}
\frac{1}{m \gamma^2}\left(\gamma F_\parallel t^\mu + F_\perp m^\mu\right)
=&&a^\mu 
+ 2v \left[t^\alpha \nabla_\alpha \eta^\mu \right]_\perp +  v t^\mu t^\alpha
  t^\rho\nabla_\rho \eta_\alpha  
\\\nonumber
&&
+ \gamma
\frac{dv}{d\tau} t^\mu + v^2 \frac{n^\mu}{R}.
\end{eqnarray}
Here $m^\mu$ is a normalized vector perpendicular to $t^\mu$ and $\eta^\mu$.
We may alternatively express \eq{kin2} in terms of the kinematical
invariants of the congruence, defined in \ref{app_kininvariant}. From
the definitions follows \cite{gravitation}%
\footnote{Note that the sign of $\omega_{\mu \nu}$ is a matter of
convention.}
\begin{eqnarray}\label{fisk}
\nabla_\nu \eta_\mu=\omega_{\mu \nu}+\theta_{\mu \nu}-a_\mu
\eta_\nu.
\end{eqnarray}
We then readily find
\begin{eqnarray}\label{hutt}
\frac{1}{m \gamma^2}\left(\gamma F_{\parallel} t^\mu +  F_{\perp} m^\mu\right)
= &&{\hspace{-0mm}}a^\mu  
+2 v \left[t^\beta ( {\omega^\mu}_\beta+{\theta^\mu}_\beta)\right]_\perp
+v t^\alpha t^\beta \theta_{\alpha \beta} t^\mu \\&&{\hspace{-0mm}} \nonumber
+ \gamma \frac{dv}{d\tau} t^\mu + v^2 \frac{n^\mu}{R}.
\end{eqnarray}
Here we have a covariant expression for the relation between spatial
projected curvature and the speed change per unit time in
terms of the experienced forces, given the kinematical invariants of
the congruence.

\subsection{Forces as experienced by the congruence observers}
It may also be interesting to know what forces are needed to
be given, by the observers following the congruence, in order to keep the
test particle on the path in question. This again is a simple
exercise of special relativity carried out in 
\ref{conforce} where we readily show that
\begin{eqnarray}
\frac{1}{\gamma^2}{P^\mu}_\alpha  \frac{D v^\alpha}{D \tau} &=&\frac{1}{\gamma m} (F_{c\parallel} t^\mu  + F_{c\perp} m^\mu).
\end{eqnarray}
Here $F_{c\parallel}$ and $F_{c\perp}$ are the experienced given
forces parallel and perpendicular to the direction of motion.  
When expressing the forces as given by
the congruence observers, it seems reasonable to express the velocity
change relative to local congruence time $d\tau_0$, given simply by
$d\tau_0=\gamma d\tau$. Then \eq{hutt} takes the form 
\begin{eqnarray}\label{hutt2}
\frac{1}{m \gamma}\left(F_{c\parallel} t^\mu + F_{c\perp} m^\mu\right)
= &&{\hspace{-0mm}}a^\mu  
+2 v \left[t^\beta ( {\omega^\mu}_\beta+{\theta^\mu}_\beta)\right]_\perp
+v t^\alpha t^\beta \theta_{\alpha \beta} t^\mu \\&&{\hspace{-0mm}} \nonumber
+ \gamma^2 \frac{dv}{d\tau_0} t^\mu + v^2 \frac{n^\mu}{R}.
\end{eqnarray}
Here we have thus the inertial force equation explicitly in terms
of the given forces. As a simple application we may consider a rotating merry-go-round with
a railway track running straight out from the center. Suppose that we
let a railway wagon move with constant speed along the track. Then
\eq{hutt2} gives us the forces on the railway track%
\footnote{After we have calculated ${\omega^\mu}_\beta$
and $a^\mu$ (for this case ${\theta^\mu}_\beta=0$). See section \ref{brat}.}.

\subsection{Discussion}\label{dis}
Looking back at \eq{hutt} and \eq{hutt2}, it is easy to put names to
the various terms. On the left hand side we have the real experienced
forces, as received and given
respectively, in the forward and sidewards direction. On the right
hand side we have
first three terms that we may call inertial forces\footnote{Actually exactly what we denote inertial force is
 subjective to a degree. For instance we could multiply all terms in
\eq{hutt} and \eq{hutt2} by $\gamma$ and define the inertial forces
accordingly.}
 given that we multiply them by $-m$:
\begin{eqnarray}
\textrm{Acceleration}:&& -m a^\mu \\
\textrm{Coriolis}:&& - 2 m v \left[t^\beta ( {\omega^\mu}_\beta+{\theta^\mu}_\beta)\right]_\perp \\
\textrm{Expansion}:&& -m v t^\alpha t^\beta \theta_{\alpha \beta} t^\mu. 
\end{eqnarray}
From a Newtonian point of view we may be tempted to call the first
inertial force 'Gravity' rather than 'Acceleration'. On the other
hand, for the particular case of using points fixed on a a rotating merry-go-round as
reference congruence, the term would correspond to what we normally
call centrifugal force. To avoid confusion we simply label this term
'Acceleration'. As regards the
second term the naming is quite obvious%
\footnote{As can be seen from \eq{fisk} (multiplied by $t^\beta$), the momentary velocities of
the congruence points (relative to an inertial system momentarily
comoving with the congruence) in the direction of motion is determined by 
$t^\beta({\omega^\mu}_\beta+{\theta^\mu}_\beta)$. Selecting the
perpendicular part gives a measure of the sideways perpendicular
velocities of the reference frame, naturally related to
Coriolis.}.
The third term is non-zero if the reference grid is expanding
or contracting in the direction of motion. 
For positive $t^\alpha t^\beta \theta_{\alpha \beta}$
the term has the form of a viscous damping force although
for negative $t^\alpha t^\beta \theta_{\alpha \beta}$ it is rather a velocity proportional driving force.
The existence of this term illustrates (for instance) that if we are using an expanding reference
frame, a real force in the direction of motion is needed to keep the
velocity relative to the reference frame fixed. 

The two last terms of \eq{hutt} and \eq{hutt2} are
\begin{eqnarray}
\gamma^2 \frac{dv}{d\tau_0} t^\mu + v^2 \frac{n^\mu}{R}.
\end{eqnarray}
These we do not regard as inertial forces, but rather as descriptions
of the motion (acceleration) relative to the reference frame. 
Notice that the formalism is well defined for arbitrary spacetimes and
arbitrary timelike congruence lines. 

\subsection{A note on alternative interpretations}
Quite commonly the term that we are here denote 'Expansion', is
included with the $dv/d\tau_0$-term (multiplied by $-m$) and these two
terms are collectively denoted the 'Euler' force. This
hides (or at least makes less manifest) the above mentioned feature that a real force is needed to
keep a fixed velocity relative to an expanding reference frame. Indeed
this lack was one of the original inspirations for making this
paper. In section \ref{reformulating} we present an alternative
approach to inertial forces resting on the notion of spatial parallel
transport (of the relativistic three-momentum relative to the
congruence). Then the expansion term arises naturally
if we are using a norm-preserving law of spatial parallel transport.

Also, quite commonly the last term, when multiplied by -$m$, in \eq{hutt} and and \eq{hutt2} respectively, is denoted the centrifugal
force. This notation is however not matching the standard
definition, where the centrifugal force comes from the acceleration due to the
rotation of the reference frame rather than from the motion of the particle
relative to the reference frame. See appendix F for further discussion of this.

If one interprets (like in e.g. \cite{ANWsta}) the two terms related to accelerations relative to
the reference frame (when multiplied by $-m$) as inertial forces
-- the whole equation takes a form of a balance equation between inertial forces. 
As interpreted in this article however, the inertial force equation is of
the standard type $F\su{real}+F\su{inertial}=m a\su{relative}$, where the acceleration
relative to the reference frame corresponds to the last two terms
of \eq{hutt} and \eq{hutt2}.

In \ref{addappendix}, we
briefly review inertial forces in Newtonian mechanics and show that
the derived formalism (and interpretation) of this paper is conforming 
(as far as that is possible) with the standard Newtonian
formalism of inertial forces, in the limit of small velocities. 
We also discuss the possibility to view the terms related to
the relative acceleration as inertial forces. For further
understanding of the viewpoint that the last two terms
are mere descriptions of the motion (acceleration) relative to the
reference frame, see also section \ref{reformulating}.

\section{A different type of curvature radius}\label{news}
In the preceding section we used somewhat different techniques in
deriving the perpendicular and the parallel parts. One might argue that the
derivation of the perpendicular part, i.e. the curvature, was in a sense
less local than the derivation of the  forward part. 

\begin{figure}[ht]
  \psfrag{Time}{Time}
  \psfrag{Spacetime trajectory}{Spacetime trajectory}
  \psfrag{Projected trajectory}{Projected trajectory}
  \psfrag{ds}{$ds$}
  \psfrag{db}{$d\bar{s}$}
  \begin{center}
        \epsfig{figure=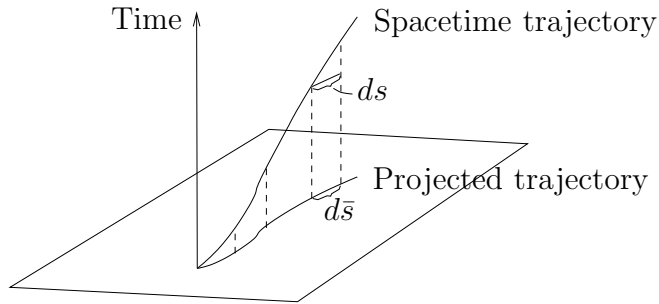,width=7cm,angle=0}
      	\caption{The difference between at-trajectory distance $ds$
        and on-slice distance $d\bar{s}$. }  
     	\label{dsdsbar}
  \end{center} 
\end{figure}

The heart of the matter lies in exactly where one measures spatial
distances. In \fig{dsdsbar} we illustrate the difference between
the on-slice distance $d\bar{s}$ and the at-trajectory distance $ds$.

To gain some intuition, we consider a finite slice, orthogonal to the congruence 
at the point where the test particle worldline intersects the slice,
and a projection of the worldline down along the
congruence onto the slice. 
The curvature radius, as defined in the preceding
section, is such that when it is infinite, the {\it on-slice} distance is
locally minimized%
\footnote{Strictly speaking it is necessary for the projected
  curvature to vanish at the point in question in order for the
  projected trajectory to minimize the distance on the slice. Note
  however that the projected curvature, as defined in the previous
  section, along the test particle worldline will not in general 
  coincide with the spatial curvature of the corresponding point along
  the projected trajectory (except at the point of intersection).}%
. 
Perhaps it would be more natural however to define a curvature
radius such that when it is infinite, the {\it at-trajectory}
distance is minimized. Obviously these two definitions will
coincide if there for instance is a Killing symmetry, and we adapt the
congruence to the Killing field. For this case $ds=d\bar{s}$, and the
two curvature measures will coincide. But in general it is perhaps not
so obvious that they will, and indeed we will find that they do not
coincide in general.

\subsection{Defining a straight line via a variational principle}
We would now like to introduce a new notion of straight trajectories,
 as those that minimize the integrated $ds$. 
We may formulate the problem of finding trajectories that are straight
in the new sense via a variational principle. We thus introduce an
action, for an arbitrary spacetime trajectory $x^\mu(\lambda)$, connecting two
fixed spacetime points
\begin{eqnarray}\label{sta}
\Delta s=\int \sqrt{P_{\mu \nu} \frac{dx^\mu}{d\lambda}
\frac{dx^\nu}{d\lambda} } d\lambda.
\end{eqnarray}
We define a corresponding Lagrangian as
\begin{eqnarray}\label{lagrange}
L=\sqrt{P_{\mu \nu} \frac{dx^\mu}{d\lambda} \frac{dx^\nu}{d\lambda} }.
\end{eqnarray}
Now we are interested in how a variation $x^\mu(\lambda)
 \rightarrow x^\mu(\lambda) + \delta x^\mu(\lambda)$ affects the
 action. Analogous (precisely) to the derivation of the Euler Lagrange
 equations we find (to first order in the change $\delta x^\mu$)
\begin{eqnarray}\label{chooses}
\delta \Delta s=\int \left[ \frac{\partial L}{\partial
x^\mu}-\frac{d}{d\lambda}\frac{\partial L}{\partial
\frac{dx^\mu}{d\lambda}} \right] \delta x^\mu d\lambda.
\end{eqnarray}
This expression holds whatever parameterization we choose. In
particular choosing the integrated local distance $s$ itself as
parameter, the Lagrangian function is unit\footnote{ So $L=1$, meaning that the
absolute derivatives of $L$ vanishes, whereas in general the partial
derivatives do not.}
along the trajectory. 
For this choice of parameter, expanding \eq{chooses} using
\eq{lagrange} is particularly simple, and the result is
\begin{eqnarray}
\delta \Delta s&=&\frac{1}{2}\int \left[ \frac{\partial P_{\alpha \beta}}{\partial
x^\mu}   \frac{dx^\alpha}{ds}\frac{dx^\beta}{ds}-2 \frac{d}{ds} \left(
P_{\mu \beta}    \frac{dx^\beta}{ds} \right) \right]  \delta x^\mu
ds\\\label{hack}
          &=&\frac{1}{2}\int \left[(\partial_\mu P_{\alpha \beta}) 
\frac{dx^\alpha}{ds} \frac{dx^\beta}{ds}
-2(\partial_\rho P_{\mu \beta})\frac{dx^\rho}{ds} \frac{dx^\beta}{ds}
-2 P_{\mu \beta} \frac{d^2 x^\beta}{ds^2}\right]  \delta x^\mu ds
\qquad.
\end{eqnarray}
Also, using $\frac{d\tau}{ds}=\frac{1}{\gamma v}$, it is easy to prove that
\begin{eqnarray}\label{change} 
\frac{d^2 x^\beta}{ds^2}=\frac{1}{\gamma v} \frac{d}{d\tau}\left(
\frac{1}{\gamma v} \frac{d x^\beta}{d\tau} \right)=...=         -\frac{1}{v^3} \frac{dv}{d\tau}
\frac{dx^\beta}{d\tau} + \frac{1}{\gamma^2 v^2} \frac{d^2 x^\beta}{d\tau^2}.
\end{eqnarray}
Inserting this into \eq{hack} we get
\begin{eqnarray}\label{hack2}
\delta \Delta s=
          \frac{1}{2}\int &&\hspace{-0mm}\left[(\partial_\mu P_{\alpha \beta}) 
\frac{dx^\alpha}{ds} \frac{dx^\beta}{ds}
-2(\partial_\rho P_{\mu \beta})\frac{dx^\rho}{ds} \frac{dx^\beta}{ds}
-2 P_{\mu \beta} \frac{1}{\gamma^2 v^2} \frac{d^2 x^\beta}{d\tau^2}
          \right. \nonumber \\
&&\left. + 2 P_{\mu \beta} \frac{1}{v^3} \frac{dv}{d\tau}
\frac{dx^\beta}{d\tau}   \right]  \delta x^\mu ds.
\end{eqnarray}
Notice that while not explicitly covariant, this expression holds (to
first order) whatever coordinates we use\footnote{This is evident since the original
equation \eq{sta}, and the derivation thus far, holds for arbitrary
coordinates.}. 
In particular it holds using
locally inertial coordinates. We can therefore change all ordinary
derivatives above to their covariant analogue \footnote{Having done
this, we may as a check-up insert the explicit expressions for the
covariant derivatives, with the affine
connection, and see that the affine connection terms indeed cancel out.}
\begin{eqnarray}\label{hack3}
\delta \Delta s=\frac{1}{2}\int &&\hspace{-0mm}\left[(\nabla_\mu P_{\alpha \beta}) 
\frac{dx^\alpha}{ds} \frac{dx^\beta}{ds}
-2(\nabla_\rho P_{\mu \beta})\frac{dx^\rho}{ds} \frac{dx^\beta}{ds}
-2 P_{\mu \beta} \frac{1}{\gamma^2 v^2} \frac{D^2 x^\beta}{D\tau^2}
          \right. \nonumber \\
&&\hspace{-0mm} \left. + 2 P_{\mu \beta} \frac{1}{v^3} \frac{dv}{d\tau}
\frac{dx^\beta}{d\tau}   \right]  \delta x^\mu ds.
\end{eqnarray}
Now we would like to see how this expression depends on $R$.
The inertial force formula \eq{rattok} can be written as 
\begin{eqnarray}
\frac{1}{\gamma^2} P_{\mu \beta} \frac{D^2 x^\beta}{D\tau^2} =&&
\eta^\alpha \nabla_\alpha \eta_\mu 
+ v \left(2 t^\alpha \nabla_\alpha \eta_\mu
-t_\mu t^\alpha t^\beta \nabla_\alpha \eta_\beta    \right)   
\\\nonumber
&&
+ \gamma^2 \frac{dv}{d\tau_0} t_\mu +
v^2 \frac{n_\mu}{R}.
\end{eqnarray}
Using this in  \eq{hack3}, together with
$\frac{dx^\alpha}{ds}=\frac{1}{v}(\eta^\alpha+v t^\alpha)$, we find after simplification
\begin{eqnarray}\label{newstraight}
\delta \Delta s=-\int  \frac{1}{v}&& \hspace{-0mm} \left[
v \frac{n_\mu}{R} 
+ t^\beta \nabla_\mu \eta_\beta 
+ \eta_\mu t^\beta a_\beta 
- t_\mu t^\alpha t^\beta \nabla_\alpha \eta_\beta \right. \\\nonumber
&&\hspace{2mm}\left.
+v \eta_\mu t^\rho t^\beta \nabla_\rho \eta_\beta
+t^\beta \nabla_\beta \eta_\mu\right]  \delta x^\mu ds.
\end{eqnarray}
This expression we may now simplify a bit. From \eq{fisk} (using the
antisymmetry of $\omega_{\mu\nu}$)
readily follows
\begin{eqnarray}
2t^\beta \theta_{\beta \mu}=
t^\beta (\nabla_\mu \eta_\beta
+ \nabla_\beta \eta_\mu)+\eta_\mu t^\beta a_\beta.
\end{eqnarray}
Using this in \eq{newstraight}, also using $t^\beta \theta_{\beta
\mu}=[t^\beta \theta_{\beta \mu}]_\perp+ t_\mu t^\alpha t^\beta
\nabla_{\alpha} \eta_\beta$, the expression within the brackets of \eq{newstraight} is
readily decomposed
into an $\eta^\mu$-part a $t^\mu$-part and a part
that is perpendicular to both $t^\mu$ and $\eta^\mu$%
\footnote{If trying to make sense, by simple thought experiments, of the
various terms -- keep in mind that while the integral
of \eq{good} corresponds to our original integral of \eq{chooses},
the {\it integrands} of these two equations are not in general the same
(recall the partial integration undertaken in the derivation of Euler
Lagrange's equations).}
\begin{eqnarray}\label{good}
\delta \Delta s=-\int\hspace{-0mm} \frac{1}{v}
\Bigg[
v \frac{n_\mu}{R} + 2[t^\alpha \theta_{\alpha \mu}]_\perp  
+\eta_\mu v t^\alpha t^\beta \nabla_\alpha \eta_\beta
+ t_\mu  t^\alpha
t^\beta \nabla_\alpha \eta_\beta   \Bigg]  \delta x^\mu ds .
\end{eqnarray}
Now, for the spacetime trajectory to be a solution to the optimization
problem, allowing for arbitrary variations $\delta x^\mu$, the
expression within the brackets must vanish. Studying the $\eta^\mu$
and $t^\mu$ parts yields
\begin{eqnarray}\label{vill}
t^\alpha t^\beta \nabla_\alpha \eta_\beta=0 .
\end{eqnarray}
What this means is that for a trajectory to optimize the integrated
distance, the trajectory must never pass two close-lying congruence
lines in a direction where there is expansion. This is actually quite
natural since there is no penalty (increase of $ds$) in letting the spacetime trajectory follow
a congruence line. To  minimize the integrated
$ds$, the spacetime trajectory must never cross between two
infinitesimally displaced congruence lines unless there is a minimum
distance separating them (implying zero expansion). It was to make
this point clear that we didn't just use the Euler Lagrange
equations directly before, but kept the expression for the change of
the action under the variation.

We are however not really interested in minimizing the distance
traveled with respect to the spacetime trajectory. In fact we just
want to minimize the integrated distance with respect to the spatial curvature.
Alternatively we could say that we want to solve the optimization problem with
respect to variations perpendicular to $t^\mu$ and $\eta^\mu$.
We see immediately from \eq{good} that this can be accomplished if
we have
\begin{eqnarray}\label{final}
v \frac{n_\mu}{R}=-2 [t^\alpha \theta_{\alpha \mu}]_\perp.
\end{eqnarray}	
We thus find that in general when there is a non-zero
expansion-shear tensor, the new sense of straightness differs from the
projected version. If we have a Killing symmetry, and
adapt the congruence to the Killing field, the congruence will
necessarily be rigid and thus $\theta_{\alpha \mu}=0$. So for a
congruence adapted to the Killing field the two curvature measures
coincide, as anticipated.
We also see that if we have isotropic expansion and no shear,
so ${\theta^\mu}_{\nu}\propto {\delta^\mu}_{\nu} $, the new sense of
straightness coincides with the projected version. This is also completely expected.

We may also notice that the projected curvature radius depends on the
velocity as $1/R \propto 1/v$. The smaller the velocity the greater
the spatial curvature (and thus the smaller the curvature {\it
radius}). This feels quite natural, moving
slowly between fixed congruence lines implies more time for expansion
and shear effects to kick in, enabling greater detours (in the
projected sense). 

\subsection{More intuition regarding the new-straight formalism}\label{more} 
\begin{figure}[ht]
  \begin{center}
    \psfrag{t}{$t$}
    \psfrag{1}{$1$}
    \psfrag{0.1}{$0.1$}
    \psfrag{y}{$y$}
    \psfrag{x}{$x$}
      	\epsfig{figure=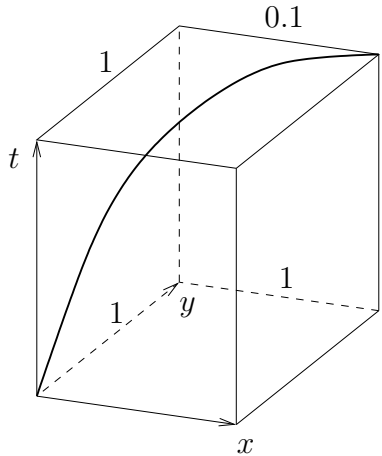,width=5cm,angle=0}
      	\caption{A box spanned by the generating congruence, seen in
      	coordinates adapted to these. The proper $x$ and $y$ distances
      	of the sides of the box are indicated. At large $t$, proper
      	distances in the $x$-direction are smaller than at small
      	$t$. We understand that a trajectory minimizing the integrated
      	proper distance will in fact curve (relative to the
      	coordinates in question), as illustrated by the thick line.}  
     	\label{box}
  \end{center} 
\end{figure}

Assume that we have a diagonal $\theta_{\alpha \beta}$ (in inertial coordinates
locally comoving with the congruence) where there is a lot of contraction
in, say the $x$-direction, and no expansion or contraction in the  $y$-direction. Consider
then, in a 2+1 
spacetime, the problem of minimizing the integrated distance
while connecting the opposing corners of a spacetime box as
illustrated in \fig{box}.

If there is severe contraction in
the $x$-direction it will towards the end of the trajectory be very cheap
to travel in the $x$-direction. Thus the trajectory should initially
start along the y-axis before turning back and at the end almost
follow the x-axis. Thus we have some intuitive understanding for why a trajectory that is straight in the new sense
has a projected curvature (in general)%
\footnote{Strictly speaking we may at least understand that such a trajectory
(as depicted in \fig{box}) {\it can} be shorter than a coordinate straight
trajectory connecting the opposite corners of the box. 
}.

\subsection{Defining a new curvature measure $\bar{R}$}
\begin{figure}[ht]
  \begin{center}
    \psfrag{k}{}
    \psfrag{dx}{\scriptsize $dx^k$}
    \psfrag{General trajectory}{General trajectory}
    \psfrag{New-straight}{New-straight}
    \psfrag{On-slice straight}{On-slice straight}
      	\epsfig{figure=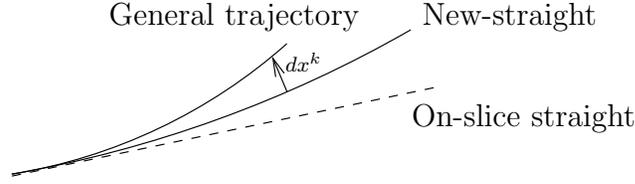,width=8cm,angle=0}
      	\caption{A projection onto the local slice of a new-straight 
        trajectory, of a
      	certain $t^\mu$ and $v$, and a corresponding projection of
      	another trajectory of the same $t^\mu$ and $v$. The deviation
      	between the lines can be used to define a new curvature and
      	curvature direction. We may understand that to lowest non-zero
      	order in time, the on-slice
      	deviation is the same as the at trajectory deviation (thinking
      	in 2+1 spacetime). So the projected deviation should do nicely
      	as a measure also of the at-trajectory deviation.}  
     	\label{curve}
  \end{center} 
\end{figure}

Now we know what kind of motion that is straight in the new
sense. Notice that the projected curvature radius of such a
line depends on both the spatial direction $t^\mu$ and the
velocity $v$. For any given $t^\mu$ and $v$ we can however define a
new curvature radius and a direction of curvature, for a general
trajectory of the $t^\mu$ and $v$ in question, by how
fast and in what direction the trajectory deviates from a
corresponding line that is straight in the new sense. See \fig{curve}.

Let $n\sut{0}^k$ and $R\sut{0}$ denote projected curvature direction and radius
respectively for the projection of a trajectory that is straight in the new sense, and let
$ds$ denote proper distance along a curve. To lowest non-zero order
the deviation is given by
\begin{eqnarray}\label{short}
dx^k=\frac{n^k}{R} \frac{ds^2}{2} -
\frac{n\sut{0}^k}{R\sut{0}} \frac{ds^2}{2}.
\end{eqnarray}
Letting a bar denote curvature direction and curvature in the new
sense, we define (see \fig{curve}) $dx^k=\frac{\bar{n}^k}{\bar{R}}\frac{ds^2}{2}$. Using
this together with \eq{final} for $n\sut{0}^k$ and $R\sut{0}$ in
\eq{short} we readily find 
\begin{eqnarray}\label{rrbar}
\frac{\bar{n}_\mu}{\bar{R}}=\frac{n_\mu}{R}+\frac{2}{v} [t^\beta \theta_{\beta \mu}]_\perp.
\end{eqnarray}
So here is the new curvature measure in terms of the projected curvature.

\subsection{The inertial force formalism using the new curvature measure}
Using \eq{rrbar} in the inertial force equation \eq{hutt}, we immediately
find the corresponding equation for the new curvature measure
\begin{eqnarray}\label{newkin}
\frac{1}{m \gamma^2} {P^\mu}_\alpha \frac{Dv^\alpha}{D\tau} = a^\mu  + 
2v t^\beta {\omega^\mu}_\beta +
v t^\alpha t^\beta \theta_{\alpha \beta} t^\mu+
\gamma^2 \frac{dv}{d\tau_0} t^\mu + v^2 \frac{\bar{n}^\mu}{\bar{R}}.
\end{eqnarray}
We see that the inertial force expression in fact is a bit cleaner with the new
representation of curvature. The difference lies in the Coriolis term,
the second term on the right hand side, which contains no shear-expansion
term now.

Notice that while we introduced the concept of curvature in the
new sense easily enough, it is a bit more abstract than the projected
curvature which can be defined via a projection onto a single locally well
defined spatial geometry. It would appear that no such geometry
applies to the new sense of curvature%
\footnote{The argument
goes like this. Suppose that we have {\it some} spatial geometry on
the local slice such that trajectories that are straight in the new
sense, and of a
certain $v$, when projected down along $\eta^\mu$ are straight relative
to the spatial geometry. Consider then trajectories with a different
velocity $v$, that are also straight in the new sense. These will (assuming a non-zero  $[t^\beta \theta_{\beta
\mu}]_\perp$) according to \eq{final} have another projected curvature
(relative to the standard spatial geometry). They will thus deviate (to
second order) from the corresponding projected trajectories of the previous
velocity. Thus these cannot also be straight relative to the spatial
geometry in question. 

We could in principle consider projecting along
some other local congruence down to a local slice such that all the
trajectories of a certain $t^\mu$ (but different $v$) that are
straight in the new sense get the same projected trajectory. To have
two effective congruences (to achieve the goal of a unique spatial
geometry) seems, at least at first sight, quite contrived and we will
not pursue the idea further here.}. 
For every fixed speed $v$, we know however what is straight in every
direction, and that is sufficient to define a curvature.

Certainly the new definition of curvature is in some sense more 'local'
than the projected one. It feels like a better match with the forward
part (connected to $dv/d\tau$) now.

\subsection{A joint expression}
For brevity it will prove useful to have a single expression that
incorporates both the projected and the new-straight formalisms. We
therefore let the suffix 's' denote either 'ps'
standing for projected straight, or 'ns' standing for
new-straight. Introducing $C\su{ps}=1$, $C\su{ns}=0$ we can then write:
\begin{eqnarray}\label{joint}
\frac{1}{m \gamma^2}\left(\gamma F_{\parallel} t^\mu +  F_{\perp} m^\mu\right)
= &&{\hspace{-0mm}}a^\mu  
+2 v \left[t^\beta ( {\omega^\mu}_\beta+C\su{s}{\theta^\mu}_\beta)\right]_\perp
+v t^\alpha t^\beta \theta_{\alpha \beta} t^\mu \\&&{\hspace{-0mm}} \nonumber
+ \gamma^2 \frac{dv}{d\tau_0} t^\mu + v^2 \frac{n\su{s}^\mu}{R\su{s}}.
\end{eqnarray}
Here $R\su{ps}\equiv R$, $R\su{ns}\equiv \bar{R}$ and analogously 
$n\su{ps}^\mu\equiv n^\mu$ and  
$n\su{ns}^\mu\equiv \bar{n}^\mu$.

\subsection{A comment on another alternative}\label{prequel}
A line that is spatially straight in the new sense is such that the
distance taken relative to the congruence is minimized (with respect
to variations in the spatial curvature). One could alternatively consider 
optimizing the {\it arrival time} for a particle
moving with a fixed speed, relative to the congruence, from one event
(along some congruence line) to another congruence line. Considering
for instance a static black hole (where there is time dilation) we
understand that to optimize the arrival time it is beneficial to
travel where there is relatively little time dilation (hence moving out
and then back relative to a straight line). We may understand that a
trajectory that is straight in the time-optimizing sense is curving
inwards relative to a line that is straight in the projected sense. We
will not pursue the issue further here, but we will comment on it
again in section \ref{hugge}.

\section{General conformal rescalings} 
In a series of papers Abramowicz et. al. (see e.g
\cite{optiskintro,marekog,Centpar,ANWsta,ANWinert,mareksurprise}) investigated
inertial forces in special and general cases using a certain conformally
rescaled spacetime. 
In this section we consider how a general
rescaling of the spacetime affects the inertial force formalism. In
section \ref{optsec} we will apply this formalism to the particular
rescaling of Abramowicz et. al.

Study then an arbitrary rescaling of spacetime
$\tilde{g}_{\mu\nu}=e^{-2\Phi} g_{\mu\nu}$. Relative to the rescaled
geometry we can express the rescaled four-acceleration of the test
particle in terms of
the rescaled curvature, the rescaled rotation tensor etc. 
Letting a tilde on an object indicate that it is related to
the rescaled spacetime, we just put tilde on
everything in the joint expression \eq{joint} for both the projected and the
new-straight formalisms. Notice that
$v$ and $\gamma$ are unaffected by the rescaling%
\footnote{One can say that space is stretched as much as time, or
that spacetime angles (and hence velocities) are preserved under
a conformal rescaling.} 
and we may omit the tilde on them
\begin{eqnarray}\label{tildejoint}
\frac{1}{\gamma^2} {\tilde{P}^\mu}{}_\alpha \frac{\tilde{D}^2 x^\alpha}{\tilde{D} \tilde{ \tau}^2}
=
&&
\tilde{a}^\mu  
+2 v \left[\tilde{t}^\beta ( {\tilde{\omega}^\mu}{}_\beta+C\su{s}{\tilde{\theta}^\mu}{}_\beta)\right]_\perp
+v\tilde{t}^\alpha \tilde{t}^\beta \tilde{\theta}_{\alpha \beta} \tilde{t}^\mu
\\\nonumber
&&
+ \gamma^2 \frac{dv}{d\tilde{\tau}_0} \tilde{t}^\mu + v^2 \frac{\tilde{n}\su{s}^\mu}{\tilde{R}\su{s}}.
\end{eqnarray}
While we have rescaled the spacetime, we are still interested in
knowing what a {\it real} observer experiences in terms of forward and
sideways thrusts. Then we need to relate the four-acceleration
relative to the rescaled spacetime to the four-acceleration of the
non-rescaled spacetime. In  \ref{kinkon} we show how the four-acceleration transforms under
conformal transformations. The result is given by \eq{kapp2}
\begin{eqnarray}\label{contrans}
\frac{\tilde{D}^2 x^\mu}{\tilde{D} \tilde{\tau}^2} =e^{2\Phi}
\frac{D^2x^\mu}{D\tau^2}  -\frac{dx^\mu}{d\tilde{\tau}}
\frac{dx^\rho}{d\tilde{\tau}} \tilde{\nabla}_\rho \Phi -\tilde{g}^{\mu \rho} \tilde{\nabla}_\rho \Phi.
\end{eqnarray}
We know that $\frac{dx^\mu}{d\tilde{\tau}}=\gamma (\tilde{\eta}^\mu +
v\tilde{t}^\mu)$ and $\frac{D^2 x^\mu}{D \tau^2}=\frac{1}{m}\left(\gamma
F_\parallel t^\mu + F_\perp m^\mu\right)$, as derived in 
\ref{app_specialforce}. Also using $\tilde{t}^\mu=e^{\Phi} t^\mu$ and
$\tilde{m}^\mu=e^{\Phi} m^\mu$, we can rewrite \eq{contrans} as
\begin{eqnarray}
\frac{1}{\gamma^2} {\tilde{P}^\mu}{}_\alpha \frac{\tilde{D}^2
x^\alpha}{\tilde{D} \tilde{ \tau}^2}=&&
\frac{1}{m\gamma^2} e^\Phi \left(\gamma
F_\parallel \tilde{t}^\mu + F_\perp \tilde{m}^\mu
\right)-
\nonumber\\
&&
\left([\tilde{P}^{\mu \rho} \tilde{\nabla}_\rho
\Phi]_\parallel+\frac{1}{\gamma^2}[\tilde{P}^{\mu \rho}
\tilde{\nabla}_\rho \Phi]_\perp + v(\tilde{\eta}^\rho\tilde{\nabla}_\rho \Phi)
\tilde{t}^\mu \right).
\end{eqnarray}
This we may now insert into \eq{tildejoint} to get an expression for the real experienced forces,
in terms of the motion relative to the rescaled spacetime, the rescaled
expansion etc. The results follow immediately. Here is the
rescaled version the inertial force expression
\begin{eqnarray}\label{much1}
\frac{e^\Phi }{m\gamma^2}\left(\gamma
F_\parallel \tilde{t}^\mu + F_\perp \tilde{m}^\mu
\right)=\tilde{a}^\mu+
[\tilde{P}^{\mu \rho} \tilde{\nabla}_\rho
\Phi]_\parallel+\frac{1}{\gamma^2}[\tilde{P}^{\mu \rho}
\tilde{\nabla}_\rho \Phi]_\perp + v(\tilde{\eta}^\rho\tilde{\nabla}_\rho \Phi)
\tilde{t}^\mu \nonumber \\
\hspace{2.0cm}+2 v \left[\tilde{t}^\beta ( {\tilde{\omega}^\mu}{}_\beta+C\su{s}{\tilde{\theta}^\mu}{}_\beta)\right]_\perp
+v\tilde{t}^\alpha \tilde{t}^\beta \tilde{\theta}_{\alpha \beta} \tilde{t}^\mu
+ \gamma^2 \frac{dv}{d\tilde{\tau}_0} \tilde{t}^\mu + v^2 \frac{\tilde{n}\su{s}^\mu}{\tilde{R}\su{s}}.
\end{eqnarray}
We notice that under general conformal rescalings, the inertial force
formalism contains extra terms, making it more
complicated in general. 

\section{Optical rescalings for a hypersurface-forming congruence}\label{optsec}
Now study the special case of a timelike hypersurface-forming congruence. The
congruence must then obey $\omega_{\mu \nu}=0$.
Such a congruence can always be generated by introducing a
foliation of spacetime specified by a single scalar function
$t(x^\mu)$. We simply form $\eta_\mu=-e^\Phi \nabla_\mu t$ (recall
that we are using the $(-,+,+,+)$-signature) where the
scalar field $\Phi$ is chosen so that $\eta^\mu$ is normalized. 

Now consider a rescaling of spacetime 
by a factor $e^{-2\Phi}$. It follows that for
displacements along the congruence we have $dt=d\tilde{\tau}$%
\footnote{Contracting both sides of $\eta_\mu=-e^\Phi \nabla_\mu t$ by
$\eta^\mu=\frac{dx^\mu}{d\tau}$ yields $1=e^\Phi \frac{dx^\mu}{d\tau} \nabla_\mu t=e^\Phi
\frac{dt}{d\tau}$. Using $d\tilde{\tau}=e^{-\Phi} d\tau$ we get $d\tilde{\tau}=dt$.}.
For this particular choice of $\Phi$ it is easy to prove, as is done
in  \ref{app_optacceleration}, that $\tilde{a}^\mu=0$. This is also easy to
understand. The rescaling, apart from stretching space, removes
time-dilation (lapse). Then it is obvious, from the point of view of maximizing
proper time, that the congruence lines are geodesics in the rescaled
spacetime. In the optically rescaled spacetime the congruence is still orthogonal
to the same slices, hence ${\tilde{\omega}}^\mu{}_\nu$ vanishes.

\subsection{The inertial force formalism in the rescaled spacetime}
Before considering the effect of the rescaling, let us for comparison
first have a look at the non-rescaled inertial force equation, for the
congruence at hand. From \eq{apphepp} in  {\ref{app_optacceleration}}, we know that
$a_\mu = {P^\alpha}_\mu \nabla_\alpha \Phi$.
Using this together with $\omega_{\alpha \beta}=0$ in \eq{joint}, we
are left with
\begin{eqnarray}\label{hipo}
\frac{1}{m \gamma^2}\left(\gamma F_\parallel t^\mu + F_\perp m^\mu\right)
=&&\hspace{-0mm}    P^{\mu \alpha} \nabla_\alpha \Phi 
+v t^\alpha t^\beta \theta_{\alpha \beta} t^\mu
+C\su{s}2 v \left[t^\beta{\theta^\mu}_\beta\right]_\perp
\nonumber\\&&\hspace{-0mm}
+ \gamma^2 \frac{dv}{d\tau_0} t^\mu + v^2 \frac{n\su{s}^\mu}{R\su{s}}.
\end{eqnarray}
Now consider a congruence and a conformal rescaling as described in the
beginning of the section.  Equation \eq{much1} is then simplified to
\begin{eqnarray}\label{much1b}
\frac{1}{m\gamma^2} e^\Phi \left(\gamma
F_\parallel \tilde{t}^\mu + F_\perp \tilde{m}^\mu
\right)=&&[\tilde{P}^{\mu \rho} \tilde{\nabla}_\rho
\Phi]_\parallel+\frac{1}{\gamma^2}[\tilde{P}^{\mu \rho}
\tilde{\nabla}_\rho \Phi]_\perp 
\\\nonumber
&&
+ v(\tilde{\eta}^\rho\tilde{\nabla}_\rho
\Phi   +  \tilde{t}^\alpha \tilde{t}^\beta \tilde{\theta}_{\alpha \beta}   )
\tilde{t}^\mu 
+C\su{s}2 v \left[\tilde{t}^\beta {\tilde{\theta}^\mu}{}_\beta \right]_\perp
\\\nonumber
&&
+ \gamma^2 \frac{dv}{d\tilde{\tau}_0} \tilde{t}^\mu + v^2 \frac{\tilde{n}\su{s}^\mu}{\tilde{R}\su{s}}.
\end{eqnarray}
While we loose the manifest connection to experienced
four-acceleration, we can further simplify this by dividing the parallel
parts of both sides by $\gamma^2$ yielding
\begin{eqnarray}\label{much1c}
\frac{1}{m\gamma^2} e^\Phi \left(
\frac{F_\parallel}{\gamma} \tilde{t}^\mu + F_\perp \tilde{m}^\mu
\right)
=
&&
\frac{1}{\gamma^2} \tilde{P}^{\mu \rho}
\tilde{\nabla}_\rho \Phi +
\frac{v}{\gamma^2}(\tilde{\eta}^\rho\tilde{\nabla}_\rho \Phi   + \tilde{t}^\alpha \tilde{t}^\beta \tilde{\theta}_{\alpha \beta}   )
\tilde{t}^\mu  \\\nonumber
&&
+C\su{s}2 v \left[\tilde{t}^\beta {\tilde{\theta}^\mu}{}_\beta \right]_\perp
+  \frac{dv}{d\tilde{\tau}_0} \tilde{t}^\mu + v^2 \frac{\tilde{n}\su{s}^\mu}{\tilde{R}\su{s}}.
\end{eqnarray}
Comparing this inertial force equation with its non-rescaled analogue
given by \eq{hipo}, we find that excepting tildes, $\gamma$-factors and
a factor $e^\Phi$, the only difference lies in the appearance of a
$\tilde{\eta}^\rho\tilde{\nabla}_\rho \Phi=\frac{\partial \Phi}{\partial t}$-term within the
expansion term. The occurrence of the extra term is quite natural considering that any
time derivative in the spacetime rescaling will act as a spatial
expansion. If we so wish, we can alternatively express \eq{much1c} in terms of
the non-rescaled kinematical invariants, see \eq{aii}-\eq{aiii} (while
still keeping the rescaled spatial curvature), thus effectively
considering a rescaled {\it space} rather than a rescaled {\it
  spacetime}.

\subsection{The projected curvature}
As a particular example we consider motion along a line that is
straight in the projected sense. The
perpendicular part of \eq{much1c} then becomes (recall that $C\su{ps}=1$)
\begin{eqnarray}
\frac{e^\Phi }{m}  F_\perp \tilde{m}^\mu
=
\tilde{P}^{\mu \rho}
\tilde{\nabla}_\rho \Phi
+ 2 v \gamma^2 \left[\tilde{t}^\beta {\tilde{\theta}^\mu}{}_\beta \right]_\perp.
\end{eqnarray}
In particular, when the rescaled congruence is rigid (so $\tilde{\theta}_{\alpha \beta}=0$), as in conformally static
spacetimes (using a suitable rescaling and a corresponding congruence) the experienced comoving
sideways force is independent of the velocity along the straight
line. This is a well known result of optical geometry. 
Now we see also how this somewhat Newtonian flavor is broken (for the
curvature measure at hand) in general when the rescaled shear-expansion tensor is non-zero.

\subsubsection{Geodesic photons}
For geodesic particles the left hand side of \eq{much1c} vanishes. In particular, for
a geodesic photon, the forward part yields simply
$\frac{dv}{d\tilde{\tau}_0}=0$, and the perpendicular part yields simply
\begin{eqnarray}
0=2 v \left[\tilde{t}^\beta {\tilde{\theta}^\mu}{}_\beta \right]_\perp+ v^2 \frac{\tilde{n}^\mu}{\tilde{R}}.
\end{eqnarray}
We see that the projected curvature vanishes for the free photon if we
have
\begin{eqnarray}\label{kol}
 {\tilde{\theta}^\mu}{}_\beta \tilde{t}^\beta \propto \tilde{t}^\mu.
\end{eqnarray}
Knowing that ${\tilde{\theta}^\mu}{}_\beta =
{\tilde{\sigma}^\mu}{}_\beta + \frac{\tilde{\theta}}{3} {\tilde{P}^\mu}{}_\beta$ we see
that \eq{kol} is equivalent to ${\tilde{\sigma}^\mu}{}_\beta
\tilde{t}^\beta \propto \tilde{t}^\mu$. We know that (may easily show that) ${\tilde{\sigma}^\mu}{}_\beta
\tilde{\eta}^\beta=0$. Knowing also that $\tilde{\sigma}_{\mu \nu}$ is a symmetric tensor it follows that in
coordinates adapted to the congruence, only the spatial part of
$\tilde{\sigma}_{\mu \nu}$ is nonzero. Also, for \eq{kol} to hold for arbitrary
spatial directions $\tilde{t}^i$, we must have $\tilde{\sigma}^i{}_j
\propto \delta^i{}_j$. Knowing also that the trace
$\tilde{\sigma}^\alpha{}_\alpha$ always vanishes, it follows that
$\tilde{\sigma}^\mu{}_\nu$ must vanish entirely. If a tensor vanishes in one
system it vanishes in all systems. Thus we conclude that for photons to
follow optical spatial geodesics in the (standard) projected meaning,
the congruence must (relative to the rescaled space) be shearfree (and
also rotationfree). It is not hard to show (see  \ref{kinkon})
that we have
$\tilde{\sigma}_{\mu \nu}=e^{-\Phi} \sigma_{\mu \nu}$ and
$\tilde{\omega}_{\mu \nu}=e^{-\Phi} \omega_{\mu \nu}$. Thus also
relative to the original spacetime geometry must the shear (and
rotation) vanish. 
This result will be used in a companion paper \cite{genopt} on
generalizing the theory of optical geometry.

\subsection{The new sense of curvature}
As a particular example we consider motion along a line that is
straight in the new sense. The
perpendicular part of \eq{much1c} then becomes
\begin{eqnarray}
 e^\Phi\frac{F_\perp}{m} \tilde{m}^\mu=
\left[ \tilde{P}^{\mu \rho} \tilde{\nabla}_\rho \Phi\right]_\perp.
\end{eqnarray}
Notice in particular the absence of $\gamma$ factors in this
expression. In a rescaled spacetime, with the new definition of
curvature, the perpendicular part works just like in Newtonian
gravity (up to a factor $e^\Phi$). The experienced sideways force is
independent of the velocity, even when the congruence is shearing
(unlike when using the projected curvature).

\subsubsection{Geodesic photons}
For geodesic particles the left hand side of \eq{much1c} vanishes. In particular, for
a geodesic photon, the forward part yields simply
$\frac{dv}{d\tilde{\tau}_0}=0$, and the perpendicular part
yields simply  
\begin{eqnarray}\label{ostraight}
\frac{\tilde{\bar{n}}^\mu}{\tilde{\bar{R}}}=0.
\end{eqnarray}
So a geodesic photon follows a line that is spatially straight in the
new sense. This results will also be used in a forthcoming paper on
generalizing the theory of optical geometry.
 
\section{Fermat's principle and its connection to straightness in
the new sense, in rescaled spacetimes}\label{hugge}
Fermat's principle (see \cite{perlick} for a formal proof) tells us
that a geodesic photon traveling from an event $P$ to a nearby
timelike trajectory $\Lambda$ will do this in such a way that the time
(as measured along $\Lambda$) is
stationary\footnote{By stationary we mean that it is a minimum or a
saddle point with respect to variations in the set of all null trajectories
connecting $P$ to $\Lambda$. As an example we may consider a 2+1 spacetime where
the spatial geometry is that of a sphere, and there is no
time-dilation. Then a geodesic photon can take the long way around
(following a great circle), rather than the short, in going from $P$
to $\Lambda$. This would be a saddle point rather than a minimum.}.
In particular any null trajectory minimizing the arrival time is a
geodesic. 

By introducing any spacelike foliation of spacetime, and a
corresponding future-increasing time coordinate $t$, optimizing the
arrival time at $\Lambda$ is equivalent to optimizing the coordinate time
difference $\delta t$. 
In particular, assuming a hypersurface-forming generating congruence,
we may introduce an orthogonal foliation and a corresponding time
coordinate $t$. After rescaling the spacetime (to take away time
dilation), coordinate time, velocity and spatial distance are related
simply by $d\tilde{s}=v dt$. The total coordinate time $\delta t$ needed
for a particle (not necessarily a photon) moving with constant speed
$v$ from $P$ to $\Lambda$ can then be expressed as
\begin{eqnarray}\label{derhuygens}
\delta t=\int dt=\int \frac{1}{v} d\tilde{s}=
\frac{1}{v} \int d\tilde{s}.
\end{eqnarray}  
What this says is quite obvious: no time dilation and constant speed
means that time is proportional to distance.
In particular for a photon, having fixed speed $v=1$, the coordinate time taken 
is minimized if and only if the integrated local (rescaled) distance is
minimized. This in turn can hold only if the curvature in the
new sense vanishes. 
So it in fact follows%
\footnote{
Strictly speaking, what we have shown is that {\it any} null geodesic
that minimizes the arrival time, for the $P$ and $\Lambda$ in question, has
vanishing $\tilde{\bar{R}}$. It seems safe to assume that any sufficiently small (but finite)
section of {\it any} null geodesic 
must correspond to minimizing the arrival time for {\it some} $P$
and $\Lambda$ (consider the equivalence principle). Since the
argumentation holds for arbitrary $P$ and $\Lambda$, it then follows that {\it any}
null geodesic has vanishing $\tilde{\bar{R}}$.}
 from Fermat's principle that a geodesic photon%
\footnote{
Logically, we have here always
referred to a photon geodesic with respect to the standard 
spacetime, which is precisely what we are after. 
We may however note that, for the particular spacetime transformation
we are considering, there is no need to
distinguish between null geodesics relative to the standard and the
rescaled spacetime. 
A null worldline is a geodesic relative to the
standard spacetime if and only if it is a geodesic relative to the
rescaled spacetime.
Indeed this follows from Fermat's principle (which in turn is very
reasonable considering the equivalence principle) since neither
null-ness, nor whether a null trajectory corresponds to a  stationary
arrival time or not, are affected by the conformal transformation.
It can also readily be shown using \eq{contrans} considering vanishing
rescaled four-acceleration, evaluating $\frac{d^2 {\bf
x}}{dt^2}$ in 
originally freely falling coordinates and then letting $\gamma
\rightarrow \infty$. 
}
 has zero curvature in the new sense relative to the
optically rescaled spacetime. 
This is a verification of our earlier result \eq{ostraight} that was 
derived without reference to Fermat's principle. 

Note also that in the rescaled spacetime any  
trajectory (not only null trajectories) of constant speed that is minimizing the spatial distance is also minimizing the arrival
time. Hence the time-optimizing curvature measure as discussed briefly
in section \ref{prequel} is identical (up to a pure rescaling) to the new-straight curvature in
the optically rescaled spacetime.

The connection between Fermat's principle, null geodesics and straight lines in the
optical geometry was realized, for conformally static spacetimes, a
long time ago. Now we see that with the new definition of curvature
the connection holds in any spacetime.

\section{Other photon related curvatures}
Besides the already discussed curvature measures, and their relation
to geodesic photons, it is not hard to come up with a couple of more
approaches with different virtues and set-backs.

\subsection{Curvature relative to that of a geodesic photon}\label{rels}
The new sense of curvature has the virtue that, in the optically rescaled
spacetime, geodesic photons follow spatially straight lines. 
On the other hand expressions like 'follow the photon' loose their
meaning in the sense that two spacetime trajectories, cutting the
same congruence lines and hence taking the 'same' spatial trajectory
(as seen in coordinates adapted to the congruence) need not have the same measure of curvature.

We could try to keep the cake, while also eating it, by using a
modified version of the projected curvature. We project the
trajectory onto the local slice, but we define the curvature -- not via the
deviation from a straight line on the slice -- but via the deviation
from the projected trajectory of a geodesic photon.  This definition
of curvature can be applied without the restriction to a 
hypersurface-forming congruence. Also, regardless of rescalings photons will per
definition follow straight lines. From \eq{hutt} it immediately
follows that a geodesic photon obeys
\begin{eqnarray}
0=\left[a^\mu\right]_\perp+2 \left[t^\beta ( {\omega^\mu}_\beta+{\theta^\mu}_\beta)\right]_\perp+\frac{n^\mu}{R}.
\end{eqnarray}
We then introduce the new curvature as
\begin{eqnarray}\label{newrett}
\frac{n'^\mu}{R'}=\frac{n^\mu}{R} + \left[a^\mu \right]_\perp +2 \left[t^\beta ( {\omega^\mu}_\beta+{\theta^\mu}_\beta)\right]_\perp.
\end{eqnarray}
Inserting this back into \eq{hutt}, and writing $a^\mu=a^\mu_\parallel + a^\mu_\perp$, yields
\begin{eqnarray}\label{finalo1}
\frac{1}{m \gamma^2}\left(\gamma F_\parallel t^\mu + F_\perp m^\mu\right)
=&&\hspace{-0mm}a^\mu_\parallel+\frac{1}{\gamma^2} a^\mu_\perp
+2v(1-v)\left[t^\beta ({\omega^\mu}_\beta+{\theta^\mu}_\beta)\right]_\perp \nonumber\\
&&\hspace{-0mm}
+v t^\alpha t^\beta \theta_{\alpha \beta} t^\mu
+ \gamma^2 \frac{dv}{d\tau_0} t^\mu + v^2 \frac{n'^\mu}{R'}.
\end{eqnarray}
Here we have then a formalism that works for an arbitrary congruence,
where geodesic photons always have zero spatial curvature -- by definition. 
If we want we can (as usual) form a single $a^\mu$-term by multiplying
the perpendicular part by $\gamma^2$. As an example, we see that for
vanishing rotation and shear, the sideways force on a particle
following a straight line (i.e. following a geodesic photon) is
independent of the velocity.

\subsection{The look-straight based curvature}\label{look}
\begin{figure}[ht]
  \begin{center}
    \psfrag{m}{}  	
    \psfrag{t}{$t^\mu$}
    \psfrag{O}{}
    \psfrag{Test-particle worldline}{Test particle worldline}
    \psfrag{Geodesic photon}{Geodesic photon}
    \epsfig{figure=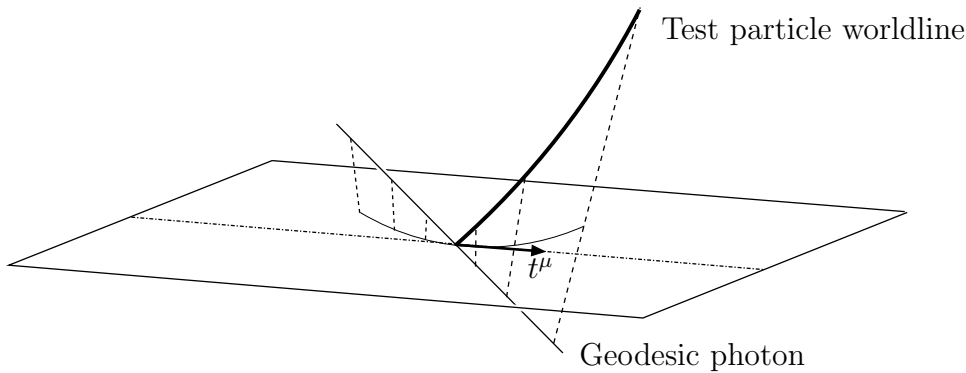,width=12cm,angle=0}
      	\caption{A 2+1 illustration, in freely falling coordinates, of a test particle following a
      	string of congruence points (dashed worldlines), momentarily seen as aligned.
	The congruence points  are those that are touched by an incoming
       	(from below in time) null geodesic in the direction
       	$-t^\mu$. In fact knowing that the upwards and downwards (in
      	time) projection of a geodesic photon passing the slice are
      	the same (as is obvious in coordinates adapted to the
      	congruence), we may understand that the projected curvature of the
      	test particle will equal the projected curvature of a geodesic
      	photon in the $-t^\mu$-direction.} 
     	\label{backphot}
  \end{center} 
\end{figure}

In \cite{Centpar} a 'seeing is
believing' principle is discussed. In a static spacetime (using the
congruence generated by the Killing field as congruence), following a
line that is {\it seen} as straight means that the experienced
comoving sideways force will be independent of the velocity. Also a
geodesic photon will follow a  trajectory that {\it looks} straight.
When there for instance is
rotation of the local reference frame we may however realize that the path taken by a geodesic photon in
fact will not look straight. We may however ask if it is possible to
define a curvature, in more general cases than the static one (using
the preferred congruence), that rests on what we {\it see}
as straight? Indeed, as is illustrated in \fig{backphot}, we
already have the necessary formalism to do this easily.
  
The figure illustrates 
that the projected curvature of a set of points that at some time was
seen as aligned in a direction $t^\mu$, in fact corresponds to the
projected curvature of a geodesic photon emitted in the
direction opposite to $t^\mu$ (i.e. the $-t^\mu$ direction).
From \eq{hutt} we immediately find (let $t^\mu\rightarrow - t^\mu$,
set $v=1$, let $F_\parallel=F_\perp=0$ and select the perpendicular part only)
\begin{eqnarray}
0=[a^\mu]_\perp  
-2 \left[t^\beta ( {\omega^\mu}_\beta+{\theta^\mu}_\beta)\right]_\perp
+\frac{n^\mu}{R}.
\end{eqnarray}
This is then the projected curvature of those congruence lines that
are momentarily {\it seen} as straight in the $t^\mu$ direction. We define a new curvature as
\begin{eqnarray}\label{newrtva}
\frac{n''^\mu}{R''}=\frac{n^\mu}{R}+\left[a^\mu\right]_\perp - 2 \left[t^\beta ( {\omega^\mu}_\beta+{\theta^\mu}_\beta)\right]_\perp.
\end{eqnarray}
Using this in \eq{hutt}, and writing $a^\mu=a^\mu_\parallel + a^\mu_\perp$, we get
\begin{eqnarray}\label{finalo2}
\frac{1}{m \gamma^2}\left(\gamma F_\parallel t^\mu + F_\perp m^\mu\right)
=&&\hspace{-0mm} a^\mu_\parallel + \frac{1}{\gamma^2} a^\mu_\perp  
+2 v (1+v)\left[t^\beta ( {\omega^\mu}_\beta+{\theta^\mu}_\beta)\right]_\perp
\nonumber \\
&&\hspace{-0mm}
+v t^\alpha t^\beta \theta_{\alpha \beta} t^\mu
+ \gamma^2 \frac{dv}{d\tau_0} t^\mu + v^2 \frac{n''^\mu}{R''}.
\end{eqnarray}
So here we have the inertial force expression when we describe our motion in
terms of what we {\it see} as straight. As always we may form a single
$a^\mu$-term if we want by dividing the parallel terms by $\gamma^2$.

We may notice that the latter definition of curvature \eq{newrtva}
matches the definition \eq{newrett} of the preceding section
(curvature relative to geodesic photon), for arbitrary $t^\mu$, if and only if
\begin{eqnarray}\label{kwpp}
\left[t^\beta ({\omega^\mu}_\beta+{\theta^\mu}_\beta)\right]_\perp=0.
\end{eqnarray}
This obviously holds when we have a rotationfree and shearfree
congruence. Also, using an argumentation similar to that under
\eq{kol}, this is also necessary for \eq{kwpp} to hold. Notice
however that \eq{kwpp} holds if we have (only) an isotropic expansion. 

\subsubsection{A comment on what looks curved}
The curvature as introduced in section \ref{look} is good measure for the curvature as
seen by a congruence observer. For the test observer that moves relative to
the congruence we must also consider beaming, making
small angular displacements from the forward direction shrink.

\subsection{General comments}
In standard optical geometry, the optical curvature radius of a
spatial line that we look upon is related to the curvature radius
that we experience by locally looking at the line, via a factor $e^\Phi$.
The latter two definitions (sections \ref{rels} and \ref{look}), for
the particular case of a static spacetime with a Killing-adapted congruence, however both correspond exactly
to the curvature that we see. 
The interesting thing with the standard optical curvature radius is
however that it is related to a global spatial geometry. Take a trajectory,
project it down onto the slice and the rescaled spatial geometry gives
us the curvature. For our two latter photon-related curvatures there
is in general (so far as I can see) no such global geometry (to which
the curvature radius is directly related), even in the static
case.
In this sense they are more abstract than the
standard optical curvature radius. On the other hand the look-straight
definition (in particular) is very operationally well defined
regardless of there being a geometry connected to it.
Lines that are seen to have a certain curvature have that very curvature, by definition. 
Actually, in this sense the standard optical curvature is not locally
well defined operationally%
\footnote{We can never figure out what $\Phi$ is through local
experiments, only its gradient can be deduced.}. 

Looking back at the joint inertial force expression \eq{much1c} for optical
rescalings utilizing the projected and the new-straight curvature
measures respectively, and comparing this with the latter two results, \eq{finalo1} and
\eq{finalo2}, we see that for a rotationfree and shearfree
congruence, a geodesic photon has zero spatial curvature in all the
four different formalisms. 
Notice however that for a rotating congruence, we cannot do the rescaling scheme (at least
not without modification), since there is no well defined slicing. 
This excludes some of the simplest and most interesting systems where
one can have use of inertial forces -- such as rotating
merry-go-rounds and stationary observers near a rotating object (like a
Kerr black hole). The latter two definitions can however be used also for these cases.

\section{Summary of the curvature measures}\label{sum}
The perpendicular part of the inertial force equation (excepting those
related to rescalings) as presented in this article is of the form
\begin{eqnarray}\label{splittocopy}
\frac{1}{\gamma^2} \left[\frac{Dv^\mu}{D\tau}\right]_\perp
= [X\su{s}^\mu]_\perp+ v^2 \frac{n\su{s}^\mu}{R\su{s}}.
\end{eqnarray}
Here the index $s$ may stand for either 'ps', 'ns', 'rp' or 'ls', corresponding
to the various curvature measures as listed in order below. For these curvature measures we
have $[X\su{s}^\mu]_\perp$ as
\begin{eqnarray}\label{s1}
\textrm{Projected Straight:} \qquad&& a^\mu_\perp + 
2v 
\left(t^\alpha{\omega^\mu}_\alpha + [t^\alpha {\theta^\mu}_\alpha]_\perp\right)  \\ 
\textrm{New-Straight:}       \qquad&& a^\mu_\perp + 2v t^\alpha {\omega^\mu}_\alpha \label{s2}\\
\textrm{Relative Photon:}    \qquad&& \frac{1}{\gamma^2} a^\mu_\perp +
2v(1-v)
\left(t^\alpha{\omega^\mu}_\alpha + [t^\alpha {\theta^\mu}_\alpha]_\perp\right) \\
\textrm{Look Straight:}\qquad&& \frac{1}{\gamma^2} a^\mu_\perp +
2v(1+v)
\left(t^\alpha{\omega^\mu}_\alpha + [t^\alpha {\theta^\mu}_\alpha]_\perp\right). \label{s4}
\end{eqnarray}
The parallel direction of the inertial force equation is given by
\begin{eqnarray}\label{forward}
\frac{1}{\gamma^2} \left[\frac{Dv^\mu}{D\tau}\right]_\parallel
= [X\su{s}^\mu]_\parallel
+\gamma \frac{dv}{d\tau} t^\mu.
\end{eqnarray}
And here
\begin{eqnarray}\label{xmupar}
[X\su{s}^\mu]_\parallel=a^\mu_\parallel + v t^\alpha t^\beta \theta_{\alpha \beta} t^\mu.
\end{eqnarray}
This part is the same for all the above curvature measures. 

Notice that all the four different physical ways of describing the motion relative to
the reference frame yield precisely the same inertial force formalisms
when using an inertial congruence (indeed there are no inertial forces
then). 

\subsection{A comment on the different ways of defining inertial forces}
As presented in this article, as is also standard for
inertial forces in Newtonian mechanics, the
final equation is of the form $F\su{real}+F\su{inertial}=m
a\su{relative}$. For a given physical scenario, the real force is
fixed, whereas the relative acceleration, and hence the inertial
forces, depend on what reference frame (congruence) we are using.
Furthermore, as we have illustrated, there is more than one plausible
way to define a spatial curvature for the motion of a test particle,
when the reference frame is shearing%
\footnote{Note incidentally that the distinction between the projected
  and the new type of curvature measure can be made also in
  non-relativistic mechanics.}. 
This effectively means that there is more than one plausible way of
  describing the acceleration relative to the reference frame -- hence
  even for a fixed reference frame there is more than one
  way of defining the inertial forces. 

As concerns the photon related approaches, they also conform to the
standard Newtonian formalism for non-shearing congruences in the limit
of small velocities. We may however note that they have somewhat of a less fundamental 
geometrical nature to them -- being more of a practical and physical nature.
Consider specifically the second photon related formalism
connected to what an observer comoving with the
reference frame in question actually experiences visually. The apparent
(inertial) forces of this formalism together with the real forces (focusing on the
perpendicular part), are precisely the (apparent) forces that will make a test
particle deviate from what the observer sees as straight.
We understand that if we let the concept of apparent (inertial) forces
be wide enough to incorporate alternative (physical) ways of measuring
the {\it apparent} motion of a test particle -- then there is room for
even more definitions of inertial forces.
Apart from the above mentioned alternative prescriptions, there is
also a certain level of freedom concerning
$\gamma$-factors, in part connected to the fact that there are two
types of forces (given and received), but see also the footnote in section \ref{dis}.

\section{Three-dimensional formalism, assuming rigid congruence}
We can rewrite the four-covariant inertial force formalism thus far as
a purely three-dimensional formalism. For brevity let us consider a
non-shearing (isotropically expanding) congruence%
\footnote{If we want to consider a shearing congruence in
three-dimensional formalism, that is in principle no problem at
all. We just define $\theta_{ij}=\frac{1}{2}(\nabla_i u_k + \nabla_k
u_i)$. Here $u^k$ is the velocity of a reference point, seen relative to a freely falling frame
locally comoving with the congruence. Also $\nabla_i$ is understood to be covariant derivative with
respect to the local spatial metric and lowering of indices are made using
the local spatial metric. The latter can be defined, as the spatial
metric on a geodesic slice (i.e. an instant in a freely falling system)
orthogonal to the congruence at a single point (the point in
question), without the existence of global orthogonal slices.
Then we could let ${\theta^\mu}_\alpha t^\alpha \rightarrow
\fat{\theta} \cdot {\bf \hat{t}}$ where $\fat{\theta}$ denotes the
three-dimensional shear-expansion matrix.}%
, so $[t^\alpha {\theta^\mu}_\alpha]_\perp=0$. Then the projected straight and the
new-straight formalisms are identical. Since $[{\theta^\mu}_{\alpha} t^\alpha]_\perp=0$ for all directions
$t^\mu$ then,  in freely falling coordinates locally comoving with the
congruence, the spatial part of ${\theta^\mu}_\nu$ must be proportional to
${\delta^i}_j$. Also, in the coordinates in question the time
components of ${\theta^\mu}_\nu$ vanishes. Defining (as is standard)
$\theta={\theta^\alpha}_\alpha$, we have thus 
${\theta^i}_j \propto \frac{\theta}{3} {\delta^i}_j$. 
We may then write
$t^\alpha t^\beta \theta_{\alpha \beta} = \frac{\theta}{3}$ (being a
scalar expression this holds in general coordinates). Looking back at
for instance \eq{hutt}, we have then
\begin{eqnarray}\label{hutte}
\frac{1}{m \gamma^2}\left(\gamma F_{\parallel} t^\mu +  F_{\perp} m^\mu\right)
= &&{\hspace{-0mm}}a^\mu  
+2 v t^\beta{\omega^\mu}_\beta
+v \frac{\theta}{3} t^\mu 
+ \gamma \frac{dv}{d\tau} t^\mu + v^2 \frac{n^\mu}{R}.
\end{eqnarray}
In coordinates locally comoving
with the congruence%
\footnote{For any specific global labeling of the congruence lines (i.e. any specific
set of spatial coordinates adapted to the congruence) we can locally
choose a time slice orthogonal to the congruence so that e.g. $a^\mu:(0,{\bf
a})$. This then uniquely defines ${\bf a}$ at any
point along the test particle trajectory.} 
we have $a^\mu:(0,{\bf a})$, $n^\mu:(0,{\bf
n})$ and and $vt^\mu:(0,{\bf v})$. 
To avoid confusion with the acceleration of the test particle,
let us define ${\bf g}=-{\bf a}$. Also we let%
\footnote{
Let $\omega^\mu=\frac{1}{2}
  \frac{1}{\sqrt{g}} \eta_\sigma \epsilon^{\sigma \mu \gamma \rho}
  \omega_{\gamma \rho}$, where $g=-\textrm{Det}[g_{\alpha
  \beta}]$ and $\epsilon^{\sigma \mu \gamma \rho}$ is $+1$, $-1$ or
  $0$ for $\sigma \mu \gamma
  \rho$  being an even, odd or no permutation of $0,1,2,3$
  respectively. Then we can define $\fat{\omega}$ through
  $\omega^\mu=(0,\fat{\omega})$ in coordinates locally
  orthogonal to the congruence.

Strictly speaking, what we mean by the cross product  ${\bf a}
\times {\bf b}$ of two three-vectors ${\bf a}$ and ${\bf b}$ is
$g^{-\frac{1}{2}}\epsilon^{ijk}a_j b_k$ where the indices have been lowered with the
local three-metric (assuming local coordinates orthogonal to the
congruence), and $g$ is minus the determinant of this metric. Notice that in general (for congruences with rotation)
there are no global time-slices that are orthogonal to the
congruence. The local three-metric corresponding to local orthogonal
coordinates is however well defined everywhere anyway. 
}
 ${\omega^\mu}_\alpha
t^\alpha \rightarrow \fat{\omega} \times {\bf \hat{t}}$
and $\tau \rightarrow \tau_0/\gamma$ (recall that
$\tau_0$ is local time along the congruence). The three-dimensional analogue of \eq{hutte} is then simply
\begin{eqnarray}\label{huttee}
\frac{1}{m \gamma^2}\left(\gamma F_{\parallel} {\bf \hat{t}} +  F_{\perp}
{\bf \hat{m}}\right)
= &&{\hspace{-0mm}} -{\bf g}  
+2 \fat{\omega} \times {\bf v} +  \frac{\theta}{3}{\bf v}
+ \gamma^2 \frac{dv}{d\tau_0} {\bf \hat{t}} + v^2 \frac{{\bf \hat{n}}}{R}.
\end{eqnarray}
Notice that this formalism is defined irrespective of whether there
exists any global slices orthogonal to the reference congruence. For
instance we can apply it to calculate the real forces on a particle
orbiting outside of the ergosphere of a Kerr black hole, using the stationary (non-rotating)
observers as our reference congruence.

Multiplying the first three terms of \eq{huttee} by $-m$ they can be seen as the
inertial forces Acceleration, Coriolis and Expansion.
The forces  $F_{\parallel}$ and $F_{\perp}$ are the experienced
(comoving) perpendicular and parallel forces respectively. If we want
to consider the {\it given} forces  $F_{c\parallel}$ and $F_{c\perp}$,
assuming that that observers following the congruence push (or pull)
the object in question, we have 
from  \ref{conforce} that
$F_{\textrm{\scriptsize c}\parallel}=F_\parallel$  and $\gamma F_{\textrm{\scriptsize c}\perp}=F_\perp$. 
Indeed defining ${\bf F}_\textrm{\scriptsize c}=F_{\textrm{\scriptsize c}\parallel}{\bf \hat{t}} + F_{\textrm{\scriptsize c}\perp}{\bf \hat{m}}$, the
inertial force equation becomes even simpler
\begin{eqnarray}\label{huttee2}
\frac{{\bf F}_\textrm{\scriptsize c}}{m \gamma}
= -{\bf g}  
+2 \fat{\omega} \times {\bf v} +  \frac{\theta}{3}{\bf v}
+ \gamma^2 \frac{dv}{d\tau_0} {\bf \hat{t}} + v^2 \frac{{\bf \hat{n}}}{R}.
\end{eqnarray}
Notice that while \eq{huttee} and \eq{huttee2} are fully
relativistically correct they are
very similar to their Newtonian counterpart(s) (just set
$\gamma=1$, see also \ref{addappendix}).
Notice however that $\tau_0$ is
local time in the reference frame. Considering for instance a static
black hole we have $d\tau_0=(1-\frac{2M}{r})^{\frac{1}{2}} dt$, where $t$ is
the global (Schwarzschild time). Also space will of course in general
be curved unlike in (standard) Newtonian theory.

\subsection{Applying the three-formalism to a rotating platform}\label{krutt}
As a simple application of the three-dimensional formalism we consider
coordinates attached to a rotating platform in special relativity. Let $\omega_0$ be
the counterclockwise angular velocity of the platform, and $r$ be the distance
from the center (this distance is obviously the same whether we are corotating with the
platform or not). We understand that the circumference of a
circle of fixed $r$ relative to the platform will (length contraction) be greater
than the corresponding circumference, as measured on the ground, by a
factor $\gamma=\gamma(\omega_0 r)$. The spatial metric in the
corotating cylindrical coordinates can thus be
written as 
\begin{eqnarray}
ds^2=dr^2 + \frac{r^2}{1-\omega_0^2 r^2/c^2} d\varphi^2+dz^2.
\end{eqnarray}
Here $c$ is the velocity of light. Note that this metric is well defined despite the fact that there are
no time slices globally orthogonal to the reference congruence in question.

For circular motion relative to an inertial
frame -- the proper acceleration, as follows from \eq{huttee}, is given by $\gamma^2 v^2/r=\gamma^2
\omega_0^2 r$. We
understand that relative to the rotating platform we have
\begin{eqnarray}\label{grot}
{\bf g}=\frac{\omega_0^2 r}{1-\omega_0^2 r^2/c^2} {\bf \hat{r}}.
\end{eqnarray}
A gyroscope orbiting with a counterclockwise angular velocity
$\omega_0$ around a circle of radius $r$ with respect to
inertial coordinates, will Thomas-precess (see e.g. \cite{formalgyro}) with a clockwise angular velocity
$\omega\su{gyro}=(\gamma-1)\omega_0$. 
Adding this rotation to
the rotation of the reference frame and multiplying by a factor
$\gamma$ to take time dilation into account, it follows that with
respect to an observer corotating with the platform, 
the gyroscope will precess with a clockwise angular velocity given by
$\gamma(\omega_0+(\gamma-1)\omega_0)=\gamma^2 \omega_0$. 
We have thus the local rotation of the platform (as experienced by a
locally comoving inertial observer)
\begin{eqnarray}\label{omrot}
\fat{\omega}=\frac{\omega_0}{1-\omega_0^2 r^2/c^2} {\bf \hat{z}}.
\end{eqnarray}
Now we have the necessary tools for making calculations with respect
to this reference frame. 

\subsection{Radial motion on the rotating platform}\label{brat}
As a particular example we consider a wagon moving along a
radially directed rail (fixed $\varphi$) on the rotating platform with constant velocity $v$. We
are interested in what force that will act on the rail from the
wagon. We note that this force is precisely minus the given force by
the rail, thus ${\bf F}\su{onrail}=-{\bf F}\su{c}$. 
Letting $m$ be the rest mass of the wagon (we assume the rotational
energy of the wheels to be negligible), we have then from \eq{huttee2}
for this simple case
\begin{eqnarray}
\frac{{\bf F}\su{onrail}}{m \gamma}
= {\bf g}  
-2 \fat{\omega} \times {\bf v}.
\end{eqnarray}
Here $\gamma=\gamma(v)$. Using \eq{grot} and \eq{omrot}, assuming the wagon to move outwards
from the center so that ${\bf v}=v {\bf \hat{r}}$, this can be written as
\begin{eqnarray}
{\bf F}\su{onrail}=m \gamma \frac{\omega_0}{1-\omega_0^2
  r^2/c^2}\left( \omega_0 r {\bf \hat{r}} -2v \fat{\hat{\varphi}}  \right).
\end{eqnarray}
Here is thus the force from the wagon on the rail. Note that the
equation applies to $r < c/\omega_0$.

\subsection{A few comments on the three-dimensional formalism}
For typical applications where the reference congruence 
lines are integral curves of 
a timelike Killing field, we can directly use \eq{hutte} and
\eq{huttee2} respectively as equations of motion, for specified
forces, to find the resulting spatial path%
\footnote{We are of course assuming that ${\bf g}$, $\fat{\omega}$ (for this case ${\bf \theta}=0$) and the
spatial geometry  are known as
functions of spatial position, 
i.e. in terms of the labeling of the congruence lines.}. 
The path can be expressed in
terms of the the test particle proper time since
$d\tau=d\tau_0/\gamma=ds/(v \gamma)$.
For the most general case however (still assuming a non-shearing reference congruence), if we want to integrate the
three-dimensional equations of motion -- we need to introduce a global
time parameter. In other words we need to introduce time slices in spacetime and associate with each
slice a parameter $t$. In general 
${\bf g}$, $\fat{\omega}$, ${\bf \theta}$ and spatial distances
between adjacent congruence lines 
will be functions of this time parameter as well as of the spatial position.
Notice however that irrespective of whether the time slices are orthogonal to
the congruence or not (in general they cannot be globally orthogonal
to the congruence) spatial distances 
are always measured proper orthogonal to the
congruence lines%
\footnote{Thus the spatial geometry is not defined as the spatial
  geometry on the slice related to the global time parameter -- consider for instance the
  example in section \ref{krutt}. Note also that the comoving
  coordinates used when introducing the bold face three-notation have
  nothing to do with the time slices related to the global time.}.
Note also that even for the stationary case, if we want to make
predictions of coincidences (like whether two particles will collide
or not) we need the global time parameter.
For a static spacetime (like a Schwarzschild black hole), 
adapting the reference congruence to the
static observers, there is a
very simple such global time $t$ where $dt=f({\bf x}) d\tau_0$ for
some function $f({\bf x})$.
Note, however, that as local equations, \eq{hutte} and \eq{huttee2} are
directly applicable, without introducing a global time, to answer
  questions like for instance what perpendicular forces one gets if
  one follows the path of a geodesic photon.

\section{A general derivation of a vector transport equation from
the inertial force formalism}\label{gener}
\begin{figure}[ht]
  \begin{center}
     \psfrag{A}{A}
     \psfrag{B}{B}
      	\epsfig{figure=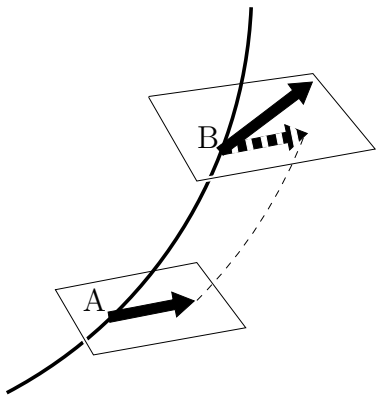,width=5cm,angle=0}
      	\caption{Vector differentiation along a timelike spacetime trajectory. 		The full
      	drawn arrows correspond to the vector defined along the
      	trajectory, for instance the momentary forward direction
      	$t^\mu$. The dashed arrow at B is the {\it transported} version of
      	the vector at A. Forming the difference between the
      	vectors at B and dividing by the proper time $d\tau$ along the
      	trajectory from A to B gives us our derivative.}  
     	\label{transport}
  \end{center} 
\end{figure}

Jantzen et. al. have also developed a covariant inertial force
formalism, see e.g
\cite{jantzen2}. They are employing various
covariant differentiations of vectors defined along a spacetime
trajectory. 
These types of covariant differentiation can readily be defined if we
have a means of transporting a vector along the trajectory in question. 
The general idea is simple, and illustrated in \fig{transport}. 

In particular one may define a spatial curvature and curvature
direction by how fast (and in what direction) the
forward direction deviates from a corresponding transported
vector. The idea is that the
transport law should somehow correspond to a spatial parallel transport
with respect to the spatial geometry defined by the congruence. That
way, the definition of spatial curvature and curvature direction is
analogous to the definition in standard Riemannian three-dimensional
differential geometry.
In the approach of this article we started from the other end by deriving the spatial curvature
measures of the various physical meanings, and we will now derive
corresponding vector transports and vector differentiations.

\subsection{Rigid congruence} \label{spat}
\begin{figure}[ht]
  \begin{center}
      	\epsfig{figure=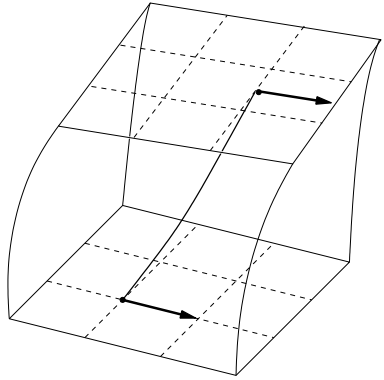,width=5cm,angle=0}
      	\caption{A 2+1 illustration of transporting a spatial vector
      	along a worldline, seen from freely falling coordinates
      	locally comoving with the congruence. As the coordinates
      	attached to the grid rotate due to ${\omega^\mu}_\alpha$, so
      	should the vector in order for it to be proper spatially
      	transported.}  
     	\label{intu}
  \end{center} 
\end{figure}

For the case of a rigid congruence%
\footnote{The congruence may rotate and accelerate but it may not shear or expand.}
the matter of spatial transport is quite intuitively reasonable. The idea is
illustrated in \fig{intu}.

It is easy to show that in the coordinates $(x^k,t)$  of a freely
 falling system, locally comoving with the congruence, the
velocity of the congruence points (assuming vanishing
${\theta^\mu}_\nu$) is to first order in $x^k$ and $t$ given by
\begin{eqnarray}
v^k={\omega^k}_j x^j + a^k t.
\end{eqnarray}
Knowing that the velocity of the congruence is zero to lowest order,
we need not worry about length contraction and such. It is then easy
to realize that the proper spacetime transport law of a vector $k^\mu$
, orthogonal to $\eta^\mu$,
corresponding to standard spatial parallel transport is
\begin{eqnarray}\label{init}
\frac{D k^\mu}{D\tau}= \gamma {\omega^\mu}_\alpha k^\alpha + b \eta^\mu.
\end{eqnarray}
Here $b$ can easily be determined from the orthogonality of $k^\mu$
and $\eta^\mu$%
\footnote{We have $k^\mu \eta_\mu=0$ which means that $\frac{D
k^\alpha}{D\tau}\eta_\alpha + k^\alpha
\frac{D\eta_\alpha}{D\tau}=0$. Contracting \eq{init} with $\eta_\mu$
then readily yields $b=k^\alpha \frac{D \eta_\alpha}{D\tau}$.}. 

\subsection{General congruence}
Now let us consider a congruence with non-zero expansion-shear tensor. Here there is no fixed
(rigid) spatial geometry. How then to define a spacetime generalization of
spatial parallel transport?

While we have no fixed spatial geometry, we still have a spatial
curvature measure (of several types) given
the spacetime trajectory. Suppose then that we transport a vector
along a timelike worldline with vanishing spatial curvature (whichever curvature measure we
choose). If the initial vector pointed in the $t^\mu$-direction it
seems natural that the parallel transported vector should keep pointing
in the $t^\mu$ direction. Also, if the trajectory curves relative to
a corresponding trajectory of vanishing spatial curvature, but the
initial vector still pointed in the $t^\mu$ direction, the transported
vector should deviate from the forward direction in the same manner as
it would for a fixed geometry.  We also demand of the parallel transport
that the norm of the vector should be constant and it should remain
orthogonal to the congruence, given that it was originally orthogonal
to the congruence. Then the derivation, as concerns parallel transport
of a vector momentarily parallel to $t^\mu$, is straightforward as we
illustrate in the coming two subsections.

\subsubsection{The standard contravariant derivative of the forward direction}
Using \eq{hrrr}, \eq{splittocopy} and \eq{fisk} we readily get
\begin{eqnarray}\label{gg1}
\left[\frac{D t^\mu}{D \tau}\right]_\perp&=&\frac{\gamma}{v}
[X\su{s}^\mu]_\perp+\gamma v \frac{n\su{s}^\mu}{R\su{s}} -
\gamma t^\alpha{\omega^\mu}_\alpha-\gamma [t^\alpha {\theta^\mu}_\alpha]_{\perp}-\frac{\gamma}{v}a_{\perp}^\mu.
\end{eqnarray}
So here is the perpendicular (spatial) part of how the forward
direction is propagated, given the spatial curvature radius%
\footnote{The $\perp$-sign on the left hand side is really only necessary for the projection
onto the slice, not to take away components in the $t^\mu$-direction
(since the normalization of $t^\mu$ tells us that there are no
$t^\mu$-component in $\frac{Dt^\mu}{D\tau}$).}. 
Notice that $X\su{s}^\mu$ depends on what curvature measure we are using (see \eq{s1}-\eq{s4}).

\subsubsection{The relation between spatial transport and spatial curvature}\label{ahhh}
Suppose now that we have some vector $t^\mu_\parallel$ that momentarily is
equal to the forward direction vector $t^\mu$. Suppose further that we
have {\it some} (as of yet undefined) parallel transport defined for
$t^\mu_\parallel$. Then we can define a curvature measure for a trajectory, with respect to the transport
in question, as
\begin{eqnarray}\label{gg2}
\gamma v \frac{n^\mu\su{v}}{R\su{v}}=
\left[\frac{Dt^\mu}{D\tau}\right]_\perp - \left[\frac{Dt^\mu_\parallel}{D\tau}\right]_\perp.
\end{eqnarray}
Here the subscript 'v', stands for 'vector transport related
curvature'. The definition is analogous to how one defines (may define)
ordinary spatial curvature using ordinary spatial parallel transport. The $\gamma$ is included since we have $\tau$ and not $\tau_0$ on the
right hand side.
Using \eq{gg1} and \eq{gg2}, making the ansatz $\frac{n^\mu\su{v}}{R\su{v}}=\frac{n\su{s}^\mu}{R\su{s}}$, hence
demanding a parallel transport of the momentarily parallel vector to
be such that the two types of curvature measures coincides, readily gives
\begin{eqnarray}
\left[\frac{Dt^\mu_\parallel}{D\tau}\right]_\perp &=&\frac{\gamma}{v}
[X\su{s}^\mu]_\perp -
\gamma t^\alpha{\omega^\mu}_\alpha-\gamma [t^\alpha {\theta^\mu}_\alpha]_{\perp}-\frac{\gamma}{v}a_{\perp}^\mu.
\end{eqnarray}
In particular for the projected and new-straight formalisms (see
\eq{s1} and \eq{s2}) this yields
\begin{eqnarray}
\textrm{Projected Straight:}
&&\left[\frac{Dt^\mu_\parallel}{D\tau}\right]_\perp = \gamma t^\alpha {\omega^\mu}_\alpha
 + \gamma [t^\alpha {\theta^\mu}_{\alpha}]_\perp \label{pstranspo}\\
\textrm{New-Straight:} &&\left[\frac{Dt^\mu_\parallel}{D\tau}\right]_\perp
=  \gamma t^\alpha {\omega^\mu}_\alpha
 - \gamma [t^\alpha {\theta^\mu}_{\alpha}]_\perp \label{nstranspo}.
\end{eqnarray}
We then define the transport equation for any vector $k^\mu_\parallel$
momentarily parallel to $t^\mu$ correspondingly
\begin{eqnarray}
\textrm{Projected Straight:}
&&\left[\frac{Dk^\mu_\parallel}{D\tau}\right]_\perp = \gamma k^\alpha_\parallel {\omega^\mu}_\alpha
 + \gamma [k^\alpha_\parallel {\theta^\mu}_{\alpha}]_\perp \label{pst}\\
\textrm{New-Straight:} &&\left[\frac{Dk^\mu_\parallel}{D\tau}\right]_\perp
=  \gamma k^\alpha_\parallel {\omega^\mu}_\alpha
 - \gamma [k^\alpha_\parallel {\theta^\mu}_{\alpha}]_\perp \label{nst}.
\end{eqnarray}
An alternative (but equivalent) way of deriving these transport laws
is to demand that the parallel
transport, along a trajectory with in general non-zero spatial
curvature, of a vector momentarily equaling the forward direction
vector should be the same as the transport of the forward direction of a
line that is {\it straight} with respect to the curvature measure in
question. 

Note that in the absence of shear, these definitions match \eq{init}. We
may however note that if we instead had considered for instance the
look-straight curvature, the corresponding transport would not have
matched \eq{init}, even for pure rotation.

\subsubsection{Spatial parallel transport of a general vector}
While the just derived transport laws are sufficient for the purposes
of the inertial force formalism, we may be curious to know whether we could find a
transport law for {\it general} vectors, that corresponds to \eq{pstranspo} and
\eq{nstranspo} for the particular case of a vector momentarily
parallel to the forward direction. Indeed we can, although how we do
it is quite subjective.

Let us however demand that, considering momentarily spatial vectors,
 the transport should be norm-preserving, and preserving orthogonality
 to $\eta^\mu$. Also we demand that any pair of parallel transported
vectors should have a fixed relative angle (in particular vectors that
were initially orthogonal should remain orthogonal). In 2+1
dimensions it is obvious, concerning spatial vectors, that these considerations completely
determine the parallel transport. In 3+1 dimensions there is however a
freedom of (spatial) rotation around the spatial direction of motion. 
Here we may however take guidance from \eq{init}, and demand that in
 the absence of shear we should get a transport corresponding to
 \eq{init}. Indeed this is not generally doable as was commented
 upon at the end of the preceding section \eq{ahhh}, although it
 will turn out to be for the case of the projected and the
 new-straight curvatures.

Let us assume that the parallel transport should be formulated in
terms of tensors, in likeliness with \eq{init}. The tensors that we
have to work with are $a^\mu$, ${\omega^\mu}_\alpha$,
${\theta^\mu}_\alpha$, $t^\mu$, $v$, $\eta^\mu$, $n^\mu$ and $R$. From these
tensors we can of course in principle form other tensors. 

To insure fixed norm and angles, the transport must effectively be a
spatial rotation relative to freely falling coordinates locally
comoving with the congruence 
\footnote{The argument is similar to that in section \ref{spat}, where length contraction will not
enter. Also there will of course be a $\eta^\mu$ term entering to
insure orthogonality.}.
Given any two orthogonal spatial vectors $d^\mu$ and $e^\mu$
we can form a rotation tensor
as $d^\mu e_\alpha -e^\mu d_\alpha$
\footnote{Forming $(d^\mu e_\alpha -e^\mu d_\alpha) k^\alpha$, for a
spatial vector $k^\alpha$, amounts to forming ${\bf d} ({\bf e} \cdot
{\bf k})-{\bf e}({\bf d} \cdot {\bf k})$ in (spatial) bold-face
notation. This is a so called vector triple product
and equals $({\bf e} \times {\bf d}) \times {\bf k}$. Thus $d^\mu
e_\alpha -e^\mu d_\alpha$ is  a rotation tensor.}. 
For brevity we define $d^\mu \wedge e_\alpha \equiv d^\mu e_\alpha -e^\mu d_\alpha$.
Several rotation tensors of this type can of course be added together to
form a net rotation tensor.

There are possibly several ways to match the above criterias but the one
we present below seems quite natural as concerns the
new-straight and the projected curvature measures. 

Looking at the
different tensors available and \eq{nstranspo} and \eq{pstranspo} it
is easy to find general transport laws, that obeys the just outlined
requirements. The spatial parts of our transport laws are given below
\begin{eqnarray}\label{hh1}
\textrm{Projected Straight:}
&&\left[\frac{Dk^\mu}{D\tau}\right]_\perp = \gamma k^\alpha {\omega^\mu}_\alpha
 + \gamma k^\alpha ([t^\beta{\theta^\mu}_\beta]_\perp \wedge t_\alpha) \\
\textrm{New-Straight:} &&\left[\frac{Dk^\mu}{D\tau}\right]_\perp
= \gamma k^\alpha {\omega^\mu}_\alpha
 - \gamma k^\alpha ( [t^\beta{\theta^\mu}_\beta]_\perp \wedge
t_\alpha) \label{hh2}.
\end{eqnarray}
Defining the transport law in such a way that a vector originally
orthogonal to the congruence remains orthogonal to the congruence, we
can add a term $\eta^\mu k^\alpha \frac{D \eta_\alpha}{D\tau}$
(analogous to what we did in section \ref{spat}) to the right hand side
of \eq{hh1} and \eq{hh2}.
That way we may remove the $\perp$ sign on the left hand side (which
was anyway there only for projection, not for orthogonality to $t^\mu$), and
express  the full transport equations.

Note that rather than $k^\alpha {\omega^\mu}_\alpha$ we might for
instance have tried $ k^\alpha (t^\beta {\omega^\mu}_\beta \wedge t_\alpha)$. These
would both give the right transport equation when $k^\mu=t^\mu$
momentarily, while in general being different for other vectors
$k^\mu$. The latter rotation version would introduce no rotation at
all around the direction of motion (the rotation vector is given by
$\fat{\omega} \times \fat{t}$, where $\fat{\omega}$ is the rotation three
vector corresponding to the rotation tensor ${\omega^\mu}_\alpha$), as seen from an inertial
system. This is however not really what we {\it want}. For a static
rotating grid it seems obvious that the parallel transport should
coincide with standard spatial parallel transport. Hence if the
congruence rotates around the direction of motion (seen from an
inertial system) so should a parallel transported vector. We should
thus use $k^\alpha {\omega^\mu}_\alpha$ rather than $ k^\alpha (t^\beta {\omega^\mu}_\beta \wedge t_\alpha)$. As regards the
${\theta^\mu}_\alpha$-term, what we want is not as clear. The way that
we have chosen gives the minimal rotation needed (seen from an
inertial system) to get the transport right. 

Note that the ambiguity in rotation around the spatial direction of motion, for
parallel transport of a general vector, has no impact on the
discussion of inertial forces. Here we are always concerned with
rotation of vectors momentarily in the forward direction, for which
case there is no ambiguity. 
The general transport laws can however be used in other contexts. In
particular one may use them when developing a relativistic three-dimensional
formalism of gyroscope precession relative to a given reference
frame. In such a formalism, see \cite{formalgyro}, the
occurrence of for instance terms of the type $\gamma k^\alpha
([t^\beta{\theta^\mu}_\beta]_\perp \wedge t_\alpha)$ follows naturally,
independent of what spatial parallel transport we consider. Thus the
form of \eq{hh1} and \eq{hh2} fits well with the formalism of
three-dimensional relativistic gyroscope precession.

\subsubsection{Covariant differentiation along trajectory}
Having derived the transport laws, the corresponding covariant
differentiations along a trajectory follows immediately. 
Including the $\eta^\mu$-component as discussed under \eq{hh2} we
simply get
\begin{eqnarray}
\frac{D\su{ps} k^\mu}{D\su{ps}\tau} &=&\frac{D k^\mu}{D \tau} -  \gamma k^\alpha {\omega^\mu}_\alpha
 - \gamma k^\alpha ([t^\beta{\theta^\mu}_\beta]_\perp \wedge t_\alpha)-\eta^\mu k^\alpha \frac{D \eta_\alpha}{D\tau} \\
\frac{D\su{ns} k^\mu}{D\su{ns}\tau}
&=&  \frac{D k^\mu}{D \tau} -\gamma k^\alpha {\omega^\mu}_\alpha
 + \gamma k^\alpha ( [t^\beta{\theta^\mu}_\beta]_\perp \wedge t_\alpha)-\eta^\mu k^\alpha \frac{D \eta_\alpha}{D\tau}.
\end{eqnarray}
Here $\frac{D \eta_\alpha}{D\tau}$ is readily given by \eq{kattaett}
and \eq{fisk}. Notice
however that for the purposes of the inertial force formalism
presented here, only
the projected part of these equations is of importance and only when
applied to a vector momentarily parallel to the forward direction.

\section{Reformulating the inertial force formalism}\label{reformulating}
Consider a rigid Cartesian reference system that rotates and possibly
accelerates in Newtonian mechanics. The law of motion can then be
expressed relative to the reference system  as (see also \ref{addappendix})
\begin{eqnarray}\label{newtin}
\frac{\fat{F}}{m}=-\frac{1}{m} \fat{F}\su{inertial} + \frac{dv}{dt} \fat{t} +
v^2 \frac{\fat{n}}{R}  .
\end{eqnarray}
Here $v$ and $R$ are the velocity and spatial curvature relative to the
reference system, analogous to the approach of the preceding
section. Alternatively we could express \eq{newtin} as
\begin{eqnarray}\label{newtin2}
\frac{\fat{F}}{m}=-\frac{1}{m}\fat{F}\su{inertial}+\frac{1}{m}\frac{d \fat{p}}{dt}.
\end{eqnarray}
Here $\fat{p}\equiv m \fat{v}$ is the three-momentum relative to
the reference system. 

The question arises if we could do something
similar in the general relativistic scheme?
Indeed we already have the necessary tools to transport
relativistic three-momentum, and do a differentiation corresponding to
$\frac{d {\bf p}}{dt}$.
When only concerned with inertial forces, there is however a more
direct way (allowing some overlap with the preceding section) as will
be presented below.

\subsection{The reformulation, with the corresponding transport in implicit form}
Let us introduce the relativistic three-momentum relative to the congruence
 as $\bar{p}^\mu \equiv {P^\mu}_\alpha p^\alpha$ (the bar here has
 nothing to do with the bar indicating new-straight curvature and curvature direction). For the particular
 case of special relativity, for an inertial congruence, \eq{rattok}
 then gives us
\begin{eqnarray}\label{rattokreduced}
\frac{1}{m \gamma^2} \frac{D\bar{p}^\mu}{D\tau}&=& \gamma \frac{dv}{d\tau} t^\mu + v^2 \frac{n^\mu}{R}.
\end{eqnarray}
By analogy with this relation we now {\it define} a covariant
differentiation of three-momentum along a curve as%
\footnote{Considering that $\frac{D \bar{p}^\mu}{D \tau}$ has an
$\eta^\mu$ component, one might think that also
  $\frac{D\su{s}
\bar{p}^\mu}{D\su{s} \tau}$ should have it. The latter is however
intended to be a differentiation between two infinitesimally different
vectors that are exactly orthogonal to $\eta^\mu$ after
some infinitesimal time. It is then easy to show that it should not
contain any explicit $\eta^\mu$ component.}
\begin{eqnarray}\label{dpdtau}
\frac{1}{m \gamma^2} \frac{D\su{s} \bar{p}^\mu}{D\su{s}\tau}&=& \gamma \frac{dv}{d\tau} t^\mu + v^2 \frac{n\su{s}^\mu}{R\su{s}}.
\end{eqnarray} 
For a general inertial force equation of the form
\begin{eqnarray}\label{kin2alt}
\frac{1}{m \gamma^2}\left(\gamma F_\parallel t^\mu + F_\perp m^\mu\right)
=X\su{s}^\mu + \gamma
\frac{dv}{d\tau} t^\mu + v^2 \frac{n\su{s}^\mu}{R\su{s}} .
\end{eqnarray}
we can thus write alternatively
\begin{eqnarray}\label{kin2alt2}
\frac{1}{m \gamma^2}\left(\gamma F_\parallel t^\mu + F_\perp m^\mu\right)
=X\su{s}^\mu +\frac{1}{m \gamma^2} \frac{D\su{s}\bar{p}^\mu}{D\su{s}\tau} .
\end{eqnarray}
Here we have then a reformulation of the inertial force formalism,
although the transport equation connected to the derivative is left implicit. We can however
derive it from the above formalism, analogous to the derivation of the
preceding section. We do this in the following section.

\subsection{Re-deriving the transport equation}
We have by definition
\begin{eqnarray}\label{defigen}
\frac{D\su{s} \bar{p}^\mu}{D\su{s}\tau}\equiv \frac{D \bar{p}^\mu}{D \tau} - \frac{D \bar{p}_\parallel^\mu}{D \tau} .
\end{eqnarray} 
Here $\bar{p}_\parallel^\mu$ is understood to be a vector that is
momentarily parallel to
$\bar{p}^\mu$ and then 'parallel' transported with respect to the
congruence (and the curvature measure in question).
This we can now use to derive the transport equation. First we write
\eq{kin2alt2} as
\begin{eqnarray}\label{kin2alt3}
\frac{1}{m \gamma^2}{P^\mu}_\alpha \frac{D p^\alpha}{D\tau}=X\su{s}^\mu +\frac{1}{m \gamma^2} \frac{D\su{s}\bar{p}^\mu}{D\su{s}\tau} .
\end{eqnarray}
Using the definitions of $\bar{p}^\mu$ and ${P^\mu}_\nu$ together
with \eq{defigen} it is then easy to show that
\begin{eqnarray}
\frac{1}{m \gamma^2}\frac{D p_\parallel^\mu}{D\tau}=X\su{s}^\mu-a^\mu-
v t^\alpha \nabla_\alpha \eta^\mu + 
\eta^\mu t^\alpha (\eta^\rho + v t^\rho) \nabla_\rho \eta_\alpha.
\end{eqnarray}
Using \eq{s1}, \eq{s2} and \eq{xmupar} together with \eq{kattaett} and \eq{fisk}
we readily find the projected version of this equation for the projected
and new-straight formalisms respectively
\begin{eqnarray}
\textrm{Projected:}\quad &&\frac{1}{m \gamma^2}{P^\mu}_\alpha \frac{D
\bar{p}_\parallel^\alpha}{D\tau}=v {\omega^\mu}_\alpha t^\alpha + v
\left[ {\theta^\mu}_\alpha t^\alpha \right]_\perp \\
\textrm{New-straight:}\quad&& \frac{1}{m \gamma^2}{P^\mu}_\alpha \frac{D
\bar{p}_\parallel^\alpha}{D\tau}=v {\omega^\mu}_\alpha t^\alpha - v
\left[ {\theta^\mu}_\alpha t^\alpha \right]_\perp.
\end{eqnarray}
These are a perfect match with \eq{pst} and \eq{nst}
(substituting $k^\mu \rightarrow \bar{p}^\alpha$). Note that for this
particular type of transport there are no ambiguities, since the vector we are
transporting is momentarily parallel to the direction of motion.

\section{The Jantzen et. al. approach revisited}
Jantzen et. al. (see e.g. \cite{jantzen3}), are using four different definitions
of covariant differentiation along curve. In the language of this
article, assuming the vector in question to be momentarily orthogonal
to $\eta^\mu$%
\footnote{If the vector has a time component we should add a term
$\gamma k^\alpha \eta_\alpha a^\mu$ on the right hand side of \eq{j3}.}%
, the definitions are%
\footnote{Note in particular that they are using a different
convention regarding the sign of ${\omega^\mu}_\alpha$, here we are
however using the convention of this article.}
\begin{eqnarray}
\frac{D\suj{fw} k^\mu}{D \tau}&=&{P^\mu}_\beta \frac{D
k^\beta}{D\tau} \label{jantztrans} \\
\frac{D\suj{cfw} k^\mu}{D \tau}&=&{P^\mu}_\beta \frac{D
k^\beta}{D\tau} - \gamma {\omega^\mu}_\alpha k^\alpha \label{j2} \\
\frac{D\suj{lie} k^\mu}{D \tau}&=&{P^\mu}_\beta \frac{D
k^\beta}{D\tau} - \gamma \left({\omega^\mu}_\alpha + {\theta^\mu}_\alpha
\right)k^\alpha \label{lie1} \label{j3}\\
\frac{D\suj{lie$\flat$} k^\mu}{D \tau}&=&{P^\mu}_\beta \frac{D
k^\beta}{D\tau} - \gamma \left({\omega^\mu}_\alpha - {\theta^\mu}_\alpha
\right)k^\alpha  \label{lie2} .
\end{eqnarray}
The subscripts are short for 'Fermi-Walker', 'Co-rotating
Fermi-Walker', 'Lie' and 'covariant Lie'. 
Note that while 'fw' really stands for Fermi-Walker \eq{jantztrans} is
not the standard Fermi-Walker derivative. 

Defining $\bar{p}^\mu \equiv m {P^\mu}_\alpha v^\alpha$, we have
$v^\mu=\gamma \eta^\mu +\frac{1}{m}\bar{p}^\mu$ and thus
\begin{eqnarray}\label{jantz}
\frac{1}{\gamma^2} {P^\mu}_\alpha \frac{D v^\alpha}{D\tau}=
\frac{1}{\gamma^2} {P^\mu}_\alpha \frac{D}{D\tau} (\gamma \eta^\alpha) +
\frac{1}{m \gamma^2} {P^\mu}_\alpha \frac{D \bar{p}^\alpha}{D \tau}.
\end{eqnarray}
We have also
\begin{eqnarray}
\frac{D\eta^\mu}{D\tau} &=&v^\alpha \nabla_\alpha \eta^\mu \\
			&=&\gamma(\eta^\alpha + v
			t^\alpha)({\theta^\mu}_\alpha+{\omega^\mu}_\alpha-a^\mu
			\eta_\alpha)\\
			&=&\gamma a^\mu + \gamma v ({\theta^\mu}_\alpha+{\omega^\mu}_\alpha) .
\end{eqnarray}
This we may use in \eq{jantz} together with \eq{jantztrans}-\eq{lie2}
(subsequently), substituting $k^\rho \rightarrow \bar{p}^\rho$. Letting 'tem'
denote 'fw','cfw','lie' or 'lie$\flat$', we immediately retrieve the
result of Jantzen et. al.
\begin{eqnarray}
\frac{1}{\gamma^2} {P^\mu}_\alpha \frac{D v^\alpha}{D\tau}= \frac{1}{m
\gamma^2} \frac{D\suj{tem} \bar{p}^\mu}{D\tau} + X^\mu\suj{tem}.
\end{eqnarray}
Here
\begin{eqnarray}
X^\mu\suj{tem}=a^\mu - v {{H\suj{tem}}^\mu}_\alpha t^\alpha.
\end{eqnarray}
Here in turn, ${{H\suj{tem}}^\mu}_\alpha$ is given by
\begin{eqnarray}
{{H\suj{fw}}^\mu}_\alpha&=&-{\omega^\mu}_\alpha - {\theta^\mu}_\alpha   \\
{{H\suj{cfw}}^\mu}_\alpha&=&-2{\omega^\mu}_\alpha - {\theta^\mu}_\alpha \\
{{H\suj{lie}}^\mu}_\alpha&=&-2{\omega^\mu}_\alpha - 2{\theta^\mu}_\alpha \label{eee1}\\
{{H\suj{lie$\flat$}}^\mu}_\alpha&=&-2{\omega^\mu}_\alpha \label{eee2}.
\end{eqnarray}
Already here we have the inertial force formalism. In the coming
subsection we will compare the two formalisms.

Jantzen et. al. has also considered an inertial force formalism in
terms of curvatures as experienced by the comoving observer \cite{jantzen4}. The idea
is essentially to study how fast, and in what direction, the incoming
congruence points are changing their velocity relative to a comoving
reference frame of gyroscopes.

\subsection{Comparing the formalisms}
We can write \eq{kin2alt3} as 
\begin{eqnarray}
\frac{1}{m \gamma^2}\frac{D v^\mu}{D\tau}=\frac{1}{m \gamma^2}
\frac{D\su{s}\bar{p}^\mu}{D\su{s}\tau} + X\su{s}^\mu.
\end{eqnarray}
Looking at \eq{s1}, \eq{s2} (we skip the relative photon and
look-straight curvature measures) and \eq{xmupar} we have $X\su{s}^\mu$ as
\begin{eqnarray}
\textrm{Projected Straight:} \quad&&X^\mu\su{ps}= a^\mu + 
2v 
\left(t^\alpha{\omega^\mu}_\alpha + [t^\alpha
{\theta^\mu}_\alpha]_\perp  \right) + v t^\alpha t^\beta \theta_{\alpha \beta} t^\mu \\ 
\textrm{New-Straight:}       \quad&&X^\mu\su{ns}= a^\mu + 2v t^\alpha
{\omega^\mu}_\alpha + v t^\alpha t^\beta \theta_{\alpha \beta} t^\mu
\end{eqnarray}
We may compare these two equations with \eq{eee1} and \eq{eee2}. We
see that as regards 
the perpendicular part, the new-straight
formalism of this article corresponds to the
$\textrm{lie}\flat$-formalism and the projected straight formalism
corresponds 
to the $\textrm{lie}$-formalism%
\footnote{The latter is expected considering the way the Lie derivative entered
the derivation of section \ref{inett}.}. The corresponding parallel parts
are however not equal. 

How one deals with the parallel part is to a large degree a matter of
taste. In this article we have defined parallel transport in such a
way that the norm of the parallel transported vector is
preserved. This is a natural definition if we want to
connect directly to changes in the local speed $v$. 
Consider for instance an isotropically expanding universe with a
particle moving along a straight line (here all the
curvature measures coincide) with constant local speed (this requires a forward
thrust) relative to the preferred congruence. With parallel transport
as defined in this article we have then
$\frac{D\bar{p}^\mu}{D\tau}=0$. In this view the forward thrust
cancels the fictitious expansion force.

The philosophy regarding the perpendicular part of the transport
equations are also a little different. We have here considered
transport equations that by definition are not altering the angles
between transported vectors, which is not generally the case in
the approach of Jantzen et. al. Again this is a matter of taste, and
it has no impact at all on the discussion of inertial forces since we 
are anyway only interested in the transport of a vector locally
aligned with the forward direction.

The biggest difference in our approaches is that we have here started
from various physically defined curvature measures, and derived an
inertial force formalism from this. Only after this was done have we
considered the notion of spatial parallel transport with respect to
the congruence. Jantzen et. al. on the other hand start from various
transport equations and derive the formalism and curvature measures
from this. 

Considering the new-straight formalism the Jantzen et. al. approach is
not really applicable. While the curvature connected to the
$lie\flat$-transport in fact corresponds to the new-straight curvature,
the connection appears coincidental. The physical meaning of this
curvature (related to minimizing the local integrated distance) has
not previously been discussed (to the author's knowledge). Neither has any
formalism previously been presented (again to the author's knowledge)
employing this curvature measure explicitly.

\section{Summary and conclusion}
The inertial force formalism as developed here was initially inspired
by the works of Abramowicz et. al. who have employed a rescaled version of
space(time) to study inertial forces. We have here extended the
formalism of inertial forces in rescaled spacetimes to include
arbitrary hypersurface-forming congruences (applicable to any
spacetime). A generalization has earlier been studied in \cite{ANWsta}
using a different philosophy, but see \cite{jantzen2} for criticism.  
We find that the inertial force formalism is very
similar in the rescaled and the standard spacetime and that the difference
lies mainly in how the $\gamma$-factors enter.
The main part of this article has turned out to be more 
connected to the work of Jantzen et. al. 
A novelty with the approach of this paper is that we are
starting from various physically defined spatial curvature measures,
and are using these to describe the local motion of a test particle,
and derive a corresponding inertial force formalism. 
In particular we
introduce a new curvature measure that we denote
new-straight curvature.  This measure is defined in such a way that,
even when we have a shearing congruence, following a straight line
with respect to the new curvature measure means taking the shortest
path relative to the spatial geometry defined by the congruence (which
is actually not the case for the standard projected curvature). This
provides us with a natural way of extending the optical
geometry, to include the most general
hypersurface-forming reference frames, while keeping the most basic
features. Indeed we show that as regards
photons, the new-straight curvature is strongly connected to Fermat's
principle. These considerations and others will be further commented
upon in a companion paper \cite{genopt} on generalizing the optical geometry.

We have also considered a pair of more unorthodox curvature measures,
the curvature relative to that of a geodesic photon and the
look-straight curvature. Likely these will have even less practical import
than the projected and the new-straight curvature measures, but they
serve as examples of the variety of different curvature measures, and
corresponding inertial force formalisms, that
one may introduce. They also illustrate how one may apply the inertial force
formalism to answer some particular questions in physics.

From the derived curvature measures, we have derived spacetime transport laws
for vectors, along a test particle worldline, corresponding to spatial
parallel transport with respect to the congruence. These transport
laws can for example be used to derive an expression for how a
gyroscope precesses relative to the reference congruence.

We have not in this paper spent much time on explaining for instance
{\it why} the sideways force increases by a $\gamma^2$-factor if we
follow a straight line in a static spacetime. For such considerations
we refer to a companion paper \cite{intu}. There we rely on simple principles such as
time dilation and the equivalence principle and derive the
relativistic three-dimensional form of the inertial force equation \eq{huttee} using no
four-covariant formalism at all.
While this paper is considerably more formal in its approach, we have
tried to employ an
(in the author's mind) more accessible mathematical notation than that
employed by Jantzen et. al.

The explicit three-formalism as presented for shearfree
(but isotropically expanding, accelerating and rotating) reference
frames is, to the author's knowledge, also novel.

\appendix

\section{The kinematical invariants of the congruence}\label{app_kininvariant}
The kinematical invariants of a congruence of worldlines of
four-velocity $\eta^\mu$ are defined
as (see e.g. \cite{gravitation})
\begin{eqnarray}
a_\mu&=&\eta^\alpha \nabla_\alpha \eta_\mu    \label{katt1} \\
\theta&=&\nabla_\alpha \eta^\alpha\\
\sigma_{\mu \nu}&=&\frac{1}{2} \left(\nabla_\rho \eta_\mu {P^\rho}_\nu +
\nabla_\rho \eta_\nu {P^\rho}_\mu \right) -\frac{1}{3} \theta P_{\mu \nu}\\
\omega_{\mu \nu}&=&\frac{1}{2} \left( \nabla_\rho \eta_\mu {P^\rho}_\nu -
\nabla_\rho \eta_\nu {P^\rho}_\mu \right) .
\end{eqnarray}
In order of appearance these objects denote the acceleration
vector, the
expansion scalar, the shear tensor and the rotation tensor. 
We will also employ what we may denote the expansion-shear tensor
\begin{eqnarray}\label{katt2}
\theta_{\mu \nu}=\frac{1}{2} \left(\nabla_\rho \eta_\mu {P^\rho}_\nu + \nabla_\rho \eta_\nu {P^\rho}_\mu \right).
\end{eqnarray}

\section{Rewriting $f^\mu$ in terms of experienced (comoving)
forward and sideways forces}\label{app_specialforce}
Consider a freely falling frame, locally comoving with $\eta^\mu$, with a
particle moving relative to this frame. In the coordinates of the inertial frame, the particle
is acted upon by a force $f^\mu$.  This force may be decomposed as
\begin{eqnarray}\label{katt}
f^\mu=f^0 \eta^\mu + f_{\parallel} t^\mu + f_{\perp} m^\mu.
\end{eqnarray}
Here $m^\mu$ is a normalized spatial vector orthogonal to $t^\mu$.

The corresponding four-force in a system locally comoving with the
particle, with velocity $v t^\mu$, is related to $f^\mu$ simply via the
Lorentz transformation. 
We may then align the first spatial coordinate with the
direction of motion, and the second with the direction of the
perpendicular force ($m^\mu$). Denoting the components of the corresponding
decomposition in the comoving system by (capital) $F$, using the fact
that $F^0=0$, the Lorentz-transformation gives us
\begin{eqnarray}
0&=&\gamma(f^0-v f_{\parallel})\\
F_{\parallel}&=&\gamma(f_{\parallel}-v f^0)\\
F_{\perp}&=&f_{\perp}.
\end{eqnarray}
From the first and second equation above follows that
$f_{\parallel}=\gamma F_{\parallel}$. Using \eq{katt},
we have then
\begin{eqnarray}\label{finapp}
{P^\mu}_\alpha f^\alpha=\gamma F_{\parallel} t^\mu + F_{\perp} m^\mu.
\end{eqnarray}
Here $F_{\parallel}$ is the experienced forward thrust (by a comoving
observer), and $F_{\perp}$  is the experienced sideways thrust. Note
that while we proved the equality in a certain system, both sides are
tensorial and thus it holds in any coordinate system.

\section{Expressing the four-acceleration in terms of the forces given
by the congruence observers}\label{conforce}
Letting $p^\mu=m v^\mu$ denote the four-momentum of a particle we have
\begin{eqnarray}
\frac{D v^\mu}{D \tau} &=& \frac{1}{m} \frac{D p^\mu}{D \tau} \\
&=&\frac{\gamma}{m} \frac{D p^\mu}{D \tau_0}.
\end{eqnarray}
Here $\tau_0$ is local time along the congruence. 
Looking at the right hand side in the coordinates of an inertial
system locally comoving with the congruence, we see that the spatial part expresses momentum transfer
per unit time i.e. force. So denoting the given forces parallel and perpendicular to
the direction of motion by $F_{\textrm{\scriptsize c}\parallel}$ and $F_{\textrm{\scriptsize c}\perp}$ we have
by definition
\begin{eqnarray}
{P^\mu}_\alpha \frac{D p^\alpha}{D \tau_0} \equiv  F_{\textrm{\scriptsize c}\parallel} t^\mu  + F_{\textrm{\scriptsize c}\perp} m^\mu.
\end{eqnarray}
Hence we have
\begin{eqnarray}
\frac{1}{\gamma^2}{P^\mu}_\alpha  \frac{D v^\alpha}{D \tau} =
\frac{1}{\gamma m} (F_{\textrm{\scriptsize c}\parallel} t^\mu  + F_{\textrm{\scriptsize c}\perp} m^\mu).
\end{eqnarray}
We may note by
comparison with \eq{finapp} that $F_{\textrm{\scriptsize c}\parallel}=F_\parallel$ and 
$F_{\textrm{\scriptsize c}\perp}=F_\perp/\gamma$.

\section{Conformal transformations, covariant differentiation and the
rescaled kinematical invariants}\label{kinkon}
Consider a conformal transformation $\tilde{g}_{\mu
\nu}=e^{-2\Phi}{g}_{\mu \nu}$. Let $k^\mu$ be a general vector field
and $\tilde{k}^\mu=e^\phi k^\mu$ its rescaled analogue.
We have then
\begin{eqnarray}\label{brrr}
\tilde{\nabla}_\mu \tilde{k}^\nu = \partial_\mu (e^\Phi k^\nu) +
\tilde{\Gamma}^\nu_{\mu \alpha} e^\Phi k^\alpha.
\end{eqnarray}
Evaluated in a system in free fall relative to the original spacetime
(so $\partial_\mu \rightarrow \nabla_\mu$), we
have
\begin{eqnarray}
 \tilde{\Gamma}^\mu_{\alpha \beta} &\stackrel{*}{=}& \frac{1}{2}
 \tilde{g}^{\mu \rho}\left(\nabla_\alpha \tilde{g}_{\rho \beta} +
 \nabla_\beta \tilde{g}_{\rho \alpha} -  \nabla_\rho \tilde{g}_{\alpha \beta}
 \right) \\
&\stackrel{*}{=}& ... \nonumber\\
&\stackrel{*}{=}&-\left[{g^\mu}_\alpha (\nabla_\beta \Phi)+ {g^\mu}_\beta (\nabla_\alpha
 \Phi) - g^{\mu \rho} (\nabla_\rho \Phi) g_{\alpha \beta} \right].
\end{eqnarray}
Using this in \eq{brrr}, evaluated in a freely falling system relative
to the original spacetime, we readily find
\begin{eqnarray}\label{con}
\tilde{\nabla}_\mu \tilde{k}^\nu=e^{\Phi}  \left( \nabla_\mu k^\nu -
{g^\nu}_\mu k^\alpha \nabla_\alpha \Phi + k_\mu g^{\nu \rho}
\nabla_\rho \Phi\right).
\end{eqnarray}
This holds in originally freely falling coordinates. Since both
sides are tensorial it holds in any coordinates.
A corresponding expression for a covariant vector
$\tilde{k}_\mu=e^{-\Phi}{k}_\mu$ is given by
\begin{eqnarray}\label{cov}
\tilde{\nabla}_\mu \tilde{k}_\nu= e^{-\Phi} \left(  \nabla_\mu k_\nu -
g_{\mu \nu} k^\alpha \nabla_\alpha \Phi + k_\mu \nabla_\nu \Phi\right).
\end{eqnarray}
Now let us apply this to the congruence invariants. The invariants are
defined according to \eq{katt1}-\eq{katt2}.
Using \eq{cov} and \eq{con}, assuming a ($-,+,+,+$) metric, we readily find the corresponding rescaled analogues
\begin{eqnarray}
\tilde{a}^\mu&=&e^{2\Phi}(a^\mu-P^{\mu \alpha}\nabla_\alpha \Phi) \label{aii}\\
\tilde{\theta}&=&e^\Phi (\theta - 3 \eta^\alpha \nabla_\alpha \Phi)\\
\tilde{\sigma}_{\mu \nu}&=&e^{-\Phi} \sigma_{\mu \nu} \\
\tilde{\omega}_{\mu \nu}&=&e^{-\Phi} \omega_{\mu \nu} \\
\tilde{\theta}_{\mu \nu}&=&e^{-\Phi} \left(\theta_{\mu \nu}-P_{\mu \nu}
\eta^\alpha \nabla_\alpha \Phi \right) \label{aiii}.
\end{eqnarray}
It may also be convenient to know how the covariant
derivative of a vector defined along a curve transforms. 
Suppose then that we have a vector $k^\mu$ defined along a trajectory of
four-velocity $v^\mu$. Let $\tilde{k}^\mu=e^\Phi k^\mu$ and
$\tilde{v}^\mu=e^\Phi v^\mu$. Considering an arbitrary smooth
extension of the vector $k^\mu$ around the worldline, we can
write%
\footnote{
We could just do an analogous derivation to that leading
to \eq{con} but for $\frac{Dk^\mu}{D\tau}$. Using the trick of
extending the vector around the trajectory we can however use the
already derived formalism and save a little time.}
$\frac{\tilde{D} \tilde{k}^\mu}{\tilde{D}\tilde{\tau}}=\tilde{v}^\alpha \tilde{\nabla}_\alpha \tilde{k}^\mu$, and apply \eq{con}
\begin{eqnarray}
\frac{\tilde{D} \tilde{k}^\mu}{\tilde{D}\tilde{\tau}}&=&\tilde{v}^\alpha
\tilde{\nabla}_\alpha \tilde{k}^\mu \\
&=& ... \nonumber\\
&=&e^{2\Phi}\left( \frac{Dk^\mu}{D\tau} -(v^\mu k^\rho -
v^\alpha k_\alpha   g^{\mu \rho}) \nabla_\rho \Phi \right) \label{dero}.
\end{eqnarray}
In particular, considering $k^\mu=v^\mu$, we get the transformation
of the four-acceleration
\begin{eqnarray}\label{fouracc}
\frac{\tilde{D} \tilde{v}^\mu}{\tilde{D}\tilde{\tau}}&=&
e^{2\Phi}\left( \frac{Dv^\mu}{D\tau} -(g^{\mu \rho} + v^\mu v^\rho) \nabla_\rho \Phi \right).
\end{eqnarray}
Equivalently we may write \eq{fouracc} 
as
\begin{equation}\label{kapp}
\begin{array}{rcl}
\label{kapp2}
\displaystyle \frac{\tilde{D}^2 x^\mu}{\tilde{D} \tilde{\tau}^2}
&=&e^{2\Phi}
\displaystyle \frac{D^2 x^\mu}{D\tau^2}  -\displaystyle \frac{dx^\mu}{d\tilde{\tau}}
\displaystyle \frac{dx^\rho}{d\tilde{\tau}} \tilde{\nabla}_\rho \Phi -\tilde{g}^{\mu
  \rho} \tilde{\nabla}_\rho \Phi
\end{array}
\end{equation}
So here is how the four-acceleration with respect to the rescaled
spacetime is related to the four-acceleration with respect to the standard spacetime.

\section{The acceleration of the generating observers in
optical geometry}\label{app_optacceleration}
We have
\begin{eqnarray}
\eta_\mu&=&-e^\Phi \nabla_\mu t  \label{t1}\\
\eta^\alpha \eta_\alpha&=&-1 \label{t2}.
\end{eqnarray}
From the normalization follows that $\eta^\alpha \nabla_\mu
\eta_\alpha=0$. 
\begin{eqnarray}
0&=&\eta^\alpha \nabla_\mu \eta_\alpha \\
 &=&-\eta^\alpha \nabla_\mu (e^\Phi \nabla_\alpha t) \\
 &=&-\eta^\alpha ( e^\Phi \nabla_\mu \Phi \nabla_\alpha t + e^\Phi \nabla_\mu
 \nabla_\alpha t) \\
 &=&-\nabla_\mu \Phi- e^\Phi \eta^\alpha \nabla_\mu
 \nabla_\alpha t.
\end{eqnarray}
This will useful when we evaluate the four-acceleration below
\begin{eqnarray}\label{apphepp}
\frac{D \eta_\mu}{D \tau}  &=&\eta^\alpha \nabla_\alpha \eta_\mu   \\
			&=&-\eta^\alpha \nabla_\alpha (e^\Phi \nabla_\mu t) \\
                        &=&-\eta^\alpha (e^\Phi \nabla_\alpha
			\Phi \nabla_\mu t +  e^\Phi \nabla_\alpha \nabla_\mu t )\\
                        &=&\eta_\mu \eta^\alpha \nabla_\alpha \Phi +\nabla_\mu \Phi \\
			&=&(\eta^\alpha \eta_\mu + {\delta^\alpha}_\mu )
			\nabla_\alpha \Phi \\
			&=&{P^\alpha}_\mu \nabla_\alpha \Phi \label{exx}.
\end{eqnarray}
We notice that the right hand side is orthogonal to $\eta^\mu$, as it must.
While the above derivation by itself had nothing to do with rescalings
of spacetime, we can still in principle consider an optically
rescaled spacetime
$\tilde{g}_{\mu \nu}=e^{-2\Phi}{g}_{\mu \nu}$,  where
$\tilde{\eta}_\mu=-\tilde{\nabla}_\mu t$. Then just
setting tildes on everything above (for the case $\Phi=0$) it
immediately follows that $\frac{\tilde{D} \tilde{\eta}^\mu}{\tilde{D}
\tilde{\tau}}=0$. This is very intuitively reasonable, because in the rescaled
spacetime there is no time dilation, thus being at
rest must maximize the proper time. Therefore, in the rescaled spacetime 
we have $\frac{\tilde{D} \tilde{\eta}^\mu}{\tilde{D}\tilde{\tau}}=0$. If we
use this, then \eq{exx} follows from \eq{aii}.

\section{A note on the Newtonian analogue}\label{addappendix}
In typical inertial force applications in the Newtonian theory, one
assumes a rigid reference frame that has an acceleration ${\bf A}_0$
of the origin and a rotation $\fat{\omega}$ that may change over time
(non-zero $\dot{\fat{\omega}}$). Following e.g. \cite{fowles} 
we have
\begin{eqnarray}\label{newton}
{\bf F}
-m{\bf A}_0
\underbrace{     -2m \fat{\omega}\times{\bf v'}   }_{{\bf F}\su{Cor}}
\underbrace{     -m\dot{\fat{\omega}}\times{\bf r'}    }_{{\bf F}\su{trans}}
\underbrace{     -m{\fat{\omega}}\times(\fat{\omega} \times {\bf r'})}_{{\bf F}\su{centrif}}
=m {\bf a'}
\end{eqnarray}
Here ${\bf F}$ is the real force and a prime means that the quantity is connected to the reference
frame in question (which is not inertial in general). In particular
${\bf a'}$ is the acceleration relative to the reference frame. From
now on we will however let ${\bf a'}={\bf a}\su{rel}$ and
drop the primes to conform with the notation of this article. 
In the above expression ${\bf F}\su{Cor}$ is the Coriolis force, ${\bf
  F}\su{centrif}$ is the centrifugal force and ${\bf F}\su{transv}$ is the transverse force.
While the former two forces have standard names, the latter does not appear to
have a universally accepted name (as pointed out in \cite{lanczos}) --
in fact in \cite{lanczos} it is denoted the Euler force.

Considering motion along a special path of local curvature direction ${\bf \hat{n}}$ and
curvature radius $R$ with respect to the reference frame, we can alternatively express the relative
acceleration as 
\begin{eqnarray}\label{accalt}
{\bf a}\su{rel}=\frac{dv}{dt} {\bf \hat{t}} +v^2 \frac{\bf \hat{n}}{R} 
\end{eqnarray}
Here ${\bf  \hat{t}}$ is the (normalized) direction of motion with
respect to the reference frame. Using \eq{accalt} the inertial force
equation then takes the form
\begin{eqnarray}\label{new2}
{\bf F}- m {\bf A}_0 + {\bf F}\su{Cor}+{\bf F}\su{transv}+{\bf F}\su{centrif}=\frac{dv}{dt} {\bf \hat{t}} +v^2 \frac{\bf \hat{n}}{R}
\end{eqnarray}
The proper (relative to an inertial frame) acceleration of a certain point ${\bf r}'$ fixed relative to the
reference frame can be found from \eq{new2}. Note that the Coriolis force and
the relative acceleration (the right hand side of \eq{new2}) are zero
for this case, thus we have 
\begin{eqnarray}
m {\bf a}\su{reference}&=& m{\bf A}_0 -{\bf F}\su{transv}-{\bf F}\su{centrif}
\end{eqnarray}
Using this in \eq{new2} and moving terms around, also using the
explicit form of the Coriolis force, we readily find
\begin{eqnarray}\label{tja}
\frac{1}{m}{\bf F}&=& {\bf a}\su{reference}+ 2 \fat{\omega}\times{\bf v}  + \frac{dv}{dt} {\bf \hat{t}} +v^2 \frac{\bf \hat{n}}{R}
\end{eqnarray}
Modulo the lack of expansion-shear terms (obviously since we are assuming
a rigid reference frame) and factors of $\gamma$, the
Newtonian formalism (in this form) precisely corresponds to the relativistic analogue \eq{hutt}, or equivalently \eq{hutt2}.

In general relativity there does not in general exist extended rigid reference
frames,  and there cannot be a general analogue of the Newtonian
version in the form of \eq{newton}. Indeed we may understand that any instance of ${\bf r'}$ should vanish for
the general case. 
Setting ${\bf r'}=0$ in \eq{newton} we note that we
reproduce \eq{tja} (since then $\bf{A}_0={\bf a}\su{ref}$). Thus, as
far as it is at all possible for general spacetimes,  \eq{hutt} and \eq{hutt2}
conform precisely with the standard Newtonian formalism.

\subsection{A two-step point of view in Newtonian mechanics}
Consider in Newtonian mechanics a rigid non-inertial reference frame. For this case we have
\begin{eqnarray}
{\bf F}\su{aparent}=\frac{dv}{dt} {\bf \hat{t}} +v^2 \frac{\bf \hat{n}}{R}
\end{eqnarray}
Here ${\bf F}\su{aparent}$ is the sum of the real and the inertial
forces. Relative to the reference frame, henceforth denoted the {\it base}
reference frame, we may choose a new reference
frame that may rotate and accelerate relative to the base reference
frame. Then we may treat ${\bf F}\su{aparent}$ just like we
treated the real force ${\bf F}$ above, to define a new frame of
reference and introduce apparent forces with respect to that frame.

In particular we note that for any particle motion relative to the base reference frame 
-- we can always, as velocity and acceleration are concerned, 
 consider the particle to momentarily
move on a circle with accelerating angular velocity. 
In a rigid coordinate system with origin at the center of the circle
 in question, and
with angular frequency and acceleration to match the particle motion, the particle is at rest and has zero acceleration
(momentarily). In these coordinates there is centrifugal force and a
transverse (Euler) force whose
magnitude and direction are given by $-mv^2\frac{\bf
  \hat{n}}{R}$ and $-m\frac{dv}{dt}{\bf \hat{t}}$ respectively. 
These two 'extra' inertial forces will then precisely balance the real
and the inertial forces as expressed relative to the base reference
frame.
Notice however that it is only in this double reference frame sense
that it makes sense denote the relative acceleration (multiplied by
$-m$) as inertial forces. Note also that by this
philosophy, for a rotating base reference frame,
we would get two types of centrifugal forces%
\footnote{Note the difference between the two forces however -- the
  first (standard) centrifugal force can be seen as a {\it field} -- living
  in the base reference frame independent of the test particle
  motion. The second is a force defined at a single point and dependent
  on the motion of the particle. Note also as concerns general
  relativity -- the very name ``centrifugal'' seems to
  indicate that there is somewhere a
 center of some relevance for the motion -- in general relativity there can naturally
 be no such a center for general spacetimes.}.

For particular applications, such as a static black hole, using a
static reference frame, the only inertial force is due to the
acceleration of the reference frame -- which one may connect in a
Newtonian sense to gravity. In standard Newtonian mechanics gravity is not an inertial
force, but a real force -- hence the original base reference frame has
a certain Newtonian inertial flavor to it (modulo curved space, time
dilation etc.). From this point of view, the extra
reference frame needed to denote the acceleration relative to
the base reference frame as
inertial forces is ``almost'' the first reference frame. 
Likely this philosophy (or something similar) is underlying the ideas
by those authors (see e.g. \cite{ANWsta,mareksurprise}) who denote the terms related to the relative
acceleration as inertial forces (when multiplied by $-m$). 
In this article we are in any case considering just a single reference frame, and
are allowing for acceleration relative to that frame.
\vspace{0.5cm}

{\bf References}
\\

\end{document}